\documentclass[a4paper, 11pt]{article}
\usepackage{comment}
\usepackage[UKenglish]{babel}
\usepackage[margin=1in]{geometry}
\usepackage[rgb]{xcolor}
\usepackage{subcaption}
\usepackage{combelow}
\usepackage{setspace}
\usepackage{csquotes}
\usepackage{microtype}
\usepackage{amsmath, amsthm, thmtools,thm-restate}
\usepackage{amssymb}
\usepackage[libertine, sb]{newtx}
\usepackage{inconsolata}
\usepackage{bm}
\usepackage{enumitem}
\usepackage[roman]{complexity}
\usepackage{footnotehyper}
\usepackage[colorlinks=true,allcolors=linkColor, hypertexnames=false]{hyperref} 
\usepackage{nameref}
\usepackage[nameinlink,capitalise]{cleveref}
\usepackage{graphicx}
\usepackage{tikz}
\usepackage{listings}

\crefname{lstlisting}{listing}{listings}
\Crefname{lstlisting}{Listing}{Listings}

\newlist{lemenumerate}{enumerate}{1}
\setlist[lemenumerate]{label=(\roman*),ref=\thelemma\,(\roman*)} %chktex 36
\crefname{lemenumeratei}{lemma}{lemmas}

\definecolor{color1}{Hsb}{100, 0.8, 0.8}
\definecolor{color2}{Hsb}{220, 0.8, 0.8}
\definecolor{color3}{Hsb}{340 , 0.8, 0.8}
\definecolor{linkColor}{Hsb}{213, 0.9, 0.80}

\declaretheorem[numberwithin=section, shaded={bgcolor=linkColor!5}]{theorem}
\declaretheorem[sibling=theorem, shaded={bgcolor=linkColor!5}]{lemma, corollary, proposition, conjecture, claim, fact}

\theoremstyle{definition}
\declaretheorem[sibling=theorem, shaded={bgcolor=linkColor!5}]{definition}

\theoremstyle{remark}
\declaretheorem[sibling=theorem]{remark, example, observation}
\makesavenoteenv{shadebox}
\Crefname{claim}{Claim}{Claims}

\renewcommand\epsilon\varepsilon
\renewcommand\emptyset\varnothing

\newcommand{\yes}{\textsc{Yes}}
\newcommand{\no}{\textsc{No}}
\newcommand{\Z}{\ensuremath{\mathbb{Z}}}

\renewcommand{\C}{\ensuremath{{\Z_5^*}}}
\renewcommand{\M}{\ensuremath{\mathcal{M}}}
\newcommand{\N}{\ensuremath{\mathcal{N}}}
\newcommand{\Cp}{\ensuremath{{\Z_p^*}}}
\renewcommand{\D}{\ensuremath{\{1,2,\frac{1}{2}\}}}
\newcommand{\Ds}{\ensuremath{\{1,2,\frac{1}{2},5\}}}
\DeclareMathOperator{\dom}{dom}
\DeclareMathOperator{\Eqn}{Eqn}
\DeclareMathOperator{\SubEqn}{SubEqn}

\DeclareMathOperator{\NAE}{NAE}
\DeclareMathOperator{\PCSP}{PCSP}
\DeclareMathOperator{\dPCSP}{Decision-PCSP}
\DeclareMathOperator{\sPCSP}{Search-PCSP}
\DeclareMathOperator{\Pol}{Pol}
\DeclareMathOperator{\CSP}{CSP}
\newcommand{\CSinglAIP}{{\mathrm{CSingl}\,\mathrm{AIP}}}
\newcommand{\kCSinglAIP}{\kappa_{\CSinglAIP}}
\newcommand{\T}{\text{1-in-3}}

\usepackage{marginnote}

\begin{document}

\title{Abelian embeddings make useful promises:\\
approximation of detrivialised equations\thanks{This work was supported by UKRI
EP/X024431/1. For the purpose of open access, the authors have applied a CC BY
public copyright licence to any Author Accepted Manuscript version arising from
this submission. All data is provided in full in the results section of this
paper.}}

\author{Tamio-Vesa Nakajima \\
Philipps-Universität Marburg\\
\texttt{nakajima@informatik.uni-marburg.de}
\and Zephyr Verwimp\\
University of Oxford\\
\texttt{zephyr.verwimp@cs.ox.ac.uk}
\and Stanislav \v{Z}ivn\'{y}\\
University of Oxford\\
\texttt{standa.zivny@cs.ox.ac.uk}
} \date{}

\maketitle
\begin{abstract}
We study the following family of search problems: given a system of equations $x+y+z\equiv 0\pmod p$ promised to admit a solution that is nonconstant on every equation, find a $c$-colouring of the variables that makes no equation monochromatic.
For $p\in\{2,3\}$, the problem recovers known tractable Promise Constraint Satisfaction Problems (PCSPs).
For $p > 3$, we obtain a complete complexity classification: the problem is polynomial-time solvable for $c \geq 4$, yet even its decision version is \NP-hard for $c = 3$, and it is not well-defined for $c = 2$.

Our main conceptual contribution is a general tractability framework for PCSPs based on Abelian embeddings, a notion introduced by Bhangale, Khot, and Minzer [STOC 2022] in the study of quantitative approximation of satisfiable CSPs. We show that suitable Abelian embeddings yield polynomial-time PCSP reductions realised by singleton algorithms in the sense of Zhuk [FOCS 2026].

On the hardness side, the promise relation is far removed from the reconfigurable regime covered by the conditional hardness framework of Banakh, Barsukov, and Nakajima [LICS 2026]; our proof instead relies on a new additive-combinatorial rigidity theorem classifying special $3$-colourings of $\Z_p$.
\end{abstract}

\section{Introduction}
In a \emph{constraint satisfaction problem} (CSP), variables take values in a fixed domain, and constraints specify which combinations of values are allowed. The task is to find an assignment that satisfies every constraint. Even when such an assignment is promised to exist, finding one can be computationally hard. This motivates studying approximation of satisfiable CSPs, where we relax the requirement either \emph{quantitatively} or \emph{qualitatively}. In quantitative approximation, we seek an assignment satisfying as many of the original constraints as possible. In qualitative approximation, we weaken the constraints and seek an assignment satisfying every weakened constraint. We illustrate this distinction with two examples of well-known \NP-hard CSPs.

Given a graph promised to be 3-colourable, a quantitative approximation seeks an assignment of three colours that gives different colours to the endpoints of as many edges as possible. A randomised polynomial-time algorithm by Goemans and Williamson~\cite{GW04} based on semidefinite programming finds a $3$-colouring that properly colours at least a $0.836$ fraction of the edges in expectation.
A qualitative approximation instead allows more colours—say, 100—while still requiring the endpoints of every edge to receive different colours. This problem was recently shown to be \NP-hard~\cite{DMR09_Journ_ConditionalHardnessApproximateColoring,GS20_Conf_DTo1HardnessColoring3ColorableGraphsO1Colors,FMW26:arxiv}.

In (monotone) 1-in-3-SAT, each clause requires exactly one of its three literals to be true. Given a satisfiable instance, quantitative approximation seeks an assignment satisfying as many clauses as possible.
A qualitative approximation can instead weaken the requirement to \emph{Not-All-Equal}: every clause must contain both a true and a false literal. This relaxation is solvable in polynomial time, as shown by Brakensiek and Guruswami~\cite{BG21}.
Many other qualitative approximations of 1-in-3-SAT have been studied, with tractability of some and \NP-hardness of others~\cite{BBB21_Conf_SymmetricPromiseConstraintSatisfactionProblemsBooleanCase,CKKNZ25_Journ_1in3VsNotAllEqualDichotomyBrokenPromise,KV26_Conf_Approximating1In3SATLinearlyOrderedHypergraph3ColouringNPHard}.
These qualitative approximation tasks are examples of \emph{Promise Constraint
Satisfaction Problems} (PCSPs)~\cite{AGH17,BG21,BBKO21_Journ_AlgebraicApproachPromiseConstraintSatisfaction}.
The contrasting complexity of these relaxations motivates studying which relations are \emph{promise-useful}~\cite{AHM2026:usefulness}: which relations admit a nontrivial relaxation such that the associated PCSP is tractable? 
Once a relation is known to be promise-useful, one can study which of its nontrivial relaxations yield tractable PCSPs.

In their series of works, Bhangale, Khot, and Minzer~\cite{BKM25_journal_CSPI,BKM25_journal_CSPII,BKM23_conf_CSPIII,BKM23_conf_CSPIV,BKM26_journal_CSPV}, and Bhangale, Khot, Liu, and Minzer~\cite{Bhangale25:focs} identified \emph{Abelian embeddings} as an important structural feature for the quantitative approximation of satisfiable CSPs. Informally, such an embedding captures additive structure within a constraint relation. Bhangale et al. use this structure in a hybrid approximation algorithm combining the solution of equations over Abelian groups with semidefinite programming~\cite{BKM26_journal_CSPV}.

\begin{definition}
    Suppose $R \subseteq A^r$ is an $r$-ary relation with $A = \dom(R)$ and $G$ is an
    Abelian group. We say that $R$ \emph{admits an Abelian embedding to
    $G$} if there exist functions $f_1, \ldots, f_r : A \to G$ such
    that (i) at least one of $f_1, \ldots, f_r$ is nonconstant, and
    (ii) for all $(x_1, \ldots, x_r) \in R$ we have
    $\sum_i f_i(x_i) = 0$. We say simply that $R$ \emph{admits an
    Abelian embedding} if it admits one to some Abelian group.
\end{definition}

In this paper, we develop an approach to qualitative approximation based on Abelian embeddings and establish a sharp complexity dichotomy for approximating detrivialised equations over prime-order groups, a family of PCSPs whose promise relations admit such embeddings. In particular, we show that suitable Abelian embeddings are a structural source of promise-usefulness.

This algorithmic role of Abelian embeddings is complemented by recent hardness results based on connectivity properties of the promise relation. Extending work of Braverman, Khot, Lifshitz, and Minzer~\cite{BKLM25_Journ_InvariancePrincipleMultisliceApplications}, Banakh, Barsukov, and Nakajima~\cite{BBN26_Conf_InfinitePCSPDichotomyMonochromaticCliques} show that, assuming the Rich 2-to-1 conjecture~\cite{Braverman21:itcs}, every (nontrivial) qualitative approximation of a symmetric reconfigurable relation is \NP-hard. Here symmetry means closure under permutations of coordinates, while reconfigurability is defined as follows.

\begin{definition}
    A relation $R$ is \emph{reconfigurable} if the graph whose vertices are tuples of $R$, and where edges are drawn between tuples which differ in exactly one coordinate, is connected.
\end{definition}

For symmetric relations, Abelian embeddability imposes a separation between tuples that cannot be overcome by local changes, while reconfigurability requires local changes to connect all tuples.
Indeed, the two regimes are disjoint: if adjacent tuples differ in coordinate $i$, then any Abelian embedding must assign the same value under $f_i$ to the two entries in that coordinate. Connectivity and symmetry therefore force every $f_i$ to be constant, so a reconfigurable relation admits no Abelian embedding.
In fact, each detrivialised equations relation defined below has an edgeless reconfiguration graph. Our hardness results therefore concern promise relations at the opposite structural extreme from the reconfigurable relations covered by the framework of Banakh, Barsukov, and Nakajima~\cite{BBN26_Conf_InfinitePCSPDichotomyMonochromaticCliques}.

Concurrent work of Chan and Ng~\cite{CN26_Prepr_UniversalAbelianEmbeddingAffineRelaxationCSP} also highlights the algorithmic role of Abelian embeddings, showing that an algorithm combining SDP and constraint singleton AIP subsumes the hybrid algorithm of Bhangale, Khot, and Minzer~\cite{BKM26_journal_CSPV}. Our tractability results give a complementary perspective for qualitative approximation (PCSPs).

\paragraph{Contributions.}
Our first contribution is a general reduction that uses suitable
Abelian embeddings and roundings to strengthen the promise relation
of a PCSP. The
reduction applies to both decision and search, yielding a new source
of tractability for qualitative approximation.

Our second contribution is a sharp complexity dichotomy for
approximating \emph{detrivialised
equations} over prime-order groups.

\begin{definition}
    Let $G$ be a finitely generated Abelian group. We define the ternary \emph{equations relation} $\Eqn(G)=\{(x,y,z)\in G^3\mid x+y+z=0\}$. Its \emph{detrivialised equations relation} is $\Eqn^*(G)=\{(x,y,z)\in\Eqn(G)\mid x,y,z\text{ are not all equal}\}$.
\end{definition}

If a ternary relation $R$ contains a constant tuple $(a,a,a)$, then every instance of $\CSP(R)$ is satisfied by assigning $a$ to every variable. Detrivialisation removes precisely these tuples: $\Eqn^*(G)$ is the largest subrelation of $\Eqn(G)$ containing no constant tuple. When $G$ has no $3$-torsion, as for $\Z_p$ with $p\neq3$, this amounts to deleting only $(0,0,0)$. For nontrivial $G$, the identity map witnesses an Abelian embedding of $\Eqn^*(G)$ into $G$.

Although systems of equations over $\Z_p$ can be solved in polynomial time,
$\CSP(\Eqn^*(\Z_p))$, i.e.~finding a solution to a system of homogeneous linear
equations mod $p$ where no particular equation is assigned a constant tuple, is
\NP-hard.\footnote{This necessarily follows from the CSP dichotomy theorem, but
rather than prove this separately we just note that this is a consequence of our
hardness results for $p \geq 5$ --- indeed, we prove below that a relaxed version of this
problem is \NP-hard. For $p = 2$ this problem is monotone 2-in-3 SAT, and
for $p = 3$ this is strong 3-colouring of 3-uniform hypergraphs, both
\NP-hard.}

We therefore study a qualitative relaxation of this problem: given a system promised to have a solution over $G$ that is nonconstant on every equation, find a $c$-colouring of its variables that makes every equation nonmonochromatic. This retains the nonconstancy requirement while discarding the equations themselves, leaving a colouring problem on the underlying hypergraph of constraints. We write $\NAE_c$ for the ternary not-all-equal relation on $c$ colours. For fixed $c$, this is the largest target relation containing no constant tuple, and hence gives the weakest relaxation excluding constant solutions.\footnote{Formally, a finite ternary relation $R$ is \emph{promise-useful}~\cite{AHM2026:usefulness} if $\PCSP(R,Q)$ is solvable in polynomial time for some finite ternary relation $Q$ with $R\to Q$ and no constant tuple.
Equivalently, $R$ is promise-useful if $\PCSP(R,\NAE_c)$ is tractable for some fixed $c\geq2$, since, after relabelling its domain, every such $Q$ is a subrelation of some $\NAE_c$.} For detrivialised equations over prime-order groups, the resulting problem is defined as follows.

\begin{definition}
    Let $p$ be a prime and $c\geq2$. An instance of \emph{$\PCSP(\Eqn^*(\Z_p), \NAE_c)$} is given by a system of linear equations of the form $x + y + z \equiv 0 \bmod p$. The goal is to output \yes{} if there exists a solution to this system over $\Z_p$ which is nonconstant on every equation, and \no{} if there is not even a solution over $\NAE_c$, i.e., a colouring of the variables with $c$ colours which renders each equation nonconstant.
\end{definition}

For $p=2$ and $c=2$, $\PCSP(\Eqn^*(\Z_p), \NAE_c)$ is (isomorphic via Boolean complementation to) the $\T$ vs. NAE-3-SAT problem, shown tractable by Brakensiek and Guruswami~\cite{BG21}. For $p=3$, $\PCSP(\Eqn^*(\Z_p), \NAE_c)$ corresponds to the $3$-rainbow vs. $c$-normal hypergraph colouring problem, studied by McDiarmid~\cite{M93_Journ_RandomRecolouringMethodGraphsHypergraphs}, who gave a randomised algorithm. A deterministic algorithm for $p\in \{2,3\}$ and all $c\geq 2$ follows from the tractability of the problem for $p=c=2$ via \emph{homomorphic relaxation}~\cite[Definition 4.6]{BBKO21_Journ_AlgebraicApproachPromiseConstraintSatisfaction}.

For $p\geq5$ however, the story is more complicated, as the following theorem shows. Note that for $p\geq 5$ the problem is well-defined for $c\geq 3$; when $c=2$, some instances satisfy both the \yes{} and \no{} conditions.

\begin{theorem}
    Fix a prime $p \geq 5$. $\PCSP(\Eqn^*(\Z_p), \NAE_c)$ is (i) solvable in polynomial time if $c \geq 4$, (ii) \NP-hard if $c = 3$.
\end{theorem}

In fact we prove a stronger theorem.
\begin{theorem}\label{general theorem}
    Fix a nontrivial finite Abelian group $G$ without elements of order 2 or 3. Then $\PCSP(\Eqn^*(G), \NAE_c)$ is (i) solvable in polynomial time if $c \geq 4$, (ii) \NP-hard if $c = 3$ (provided the problem is well-defined).
\end{theorem}

The algorithm that solves this problem is a \emph{singleton} algorithm~\cite{Zhuk26:focs}, and its correctness relies on the fact that $\Eqn^*(G)$ has an Abelian embedding to $G$. In fact, the algorithm generalises easily to one which works for all templates $\PCSP(R,\NAE_c)$, where $R$ has an Abelian embedding to a group with a ``$c$-rounding''. We present this more general algorithm along with the notion of $c$-rounding in~\Cref{sec:tractability}, and show that the tractability part of the theorem above is a simple consequence.
However, neither $k$-consistency (for any fixed $k$) nor the combined Basic Linear Programming and Affine Integer Programming (BLP+AIP) relaxation~\cite{BGWZ20_Journ_PowerCombinedBasicLPAffineRelaxationPromiseCSPs} solves this problem (\Cref{appendix:failure-k-cons-BLPAIP}). 
Thus this is an example of a natural fragment of finite-domain PCSPs solvable by a singleton algorithm but not by either of the above methods.

\paragraph{Algorithmic results.}
We illustrate the algorithm by running it on a small instance of
$\PCSP(\Eqn^*(\Z_5),\NAE_4)$. In fact, we will run it on the search version of the problem, denoted by $\sPCSP(\Eqn^*(\Z_5),\NAE_4)$: Given an instance that is promised to admit a solution over $\Eqn^*(\Z_5)$, the goal is to find a solution over $\NAE_4$.
Consider the instance 
with variable set
$\{a,b,c,d,e,f,g,x,y,z,u\}$ and the following eight equation constraints,
where all equations are modulo~$5$:
\[
    \begin{gathered}
        C_1:a+b+e=0,\quad C_2:a+c+d=0,\quad
        C_3:a+d+g=0,\quad C_4:a+e+f=0,\\
        C_5:b+c+g=0,\quad C_6:b+d+f=0,\quad
        C_7:x+y+z=0,\quad C_8:a+x+u=0.
    \end{gathered}
\]
This is a valid instance: the assignment
\[
    \begin{gathered}
        a=0,\quad b=1,\quad c=2,\quad d=3,\quad e=4,\quad f=1,\quad
        g=2,\quad x=1,\quad y=1,\quad z=3,\quad u=4
    \end{gathered}
\]
satisfies every equation nontrivially. The algorithm is
not given the displayed assignment; it only uses the promise that such an
assignment exists. We will now explain how the algorithm works (illustrated in \Cref{fig:peeling-example}).

We start by picking some constraint, say $C_1$. We know there exists some
solution to the system such that $C_1$ does not get assigned the tuple
$(0,0,0)$. We find such a solution as follows. We choose a variable in $C_1$,
say $a$, and look (using Gaussian elimination) for a solution to the system with
$a$ fixed to $1$ or some other nonzero value. In this case, no such solution
exists, so we move on to another variable in $C_1$, say $b$, and look for a
solution to the system with $b$ fixed to $1$. We find such a solution $s_1$, and
fix the values of all variables that got a nonzero value. Next, we repeat the
same process for the subinstance induced on the remaining variables. The final
$\NAE_4$-solution $\gamma$ assigns each variable the first nonzero value it
receives in \(s_1,s_2,s_3\). The \textit{end} row of the lower table in~\Cref{fig:peeling-example} lists the final tuple assigned to each
constraint and shows that every constraint is nonmonochromatic. Informally, the
reason this works is that, in the first step at which any variable in a
constraint is coloured at all, it must be the case that 2 or 3 variables in that
constraint are coloured, and indeed coloured with at least 2 different values.
Hence, no matter what colour is assigned to the remaining variable (if it exists), the constraint is satisfied.

\begin{figure}[htb]
\begin{center}
\begin{minipage}{\textwidth}
    \centering
    \newcommand{\fixedvalue}[1]{%
        \tikz[baseline=(value.base)]
        \node[draw=linkColor,fill=linkColor!10,minimum size=1.45em,
              inner sep=0pt,rounded corners=1pt] (value) {$#1$};}
    \newcommand{\zerovalue}{%
        \tikz[baseline=(value.base)]
        \node[draw=black!45,dashed,fill=black!4,minimum size=1.45em,
              inner sep=0pt,rounded corners=1pt] (value) {$0$};}
    \newcommand{\inactivevalue}{\textcolor{black!30}{--}}
    \newcommand{\certvalue}[1]{%
        \tikz[baseline=(value.center)]
        \node[draw=linkColor,fill=linkColor!10,minimum size=1.45em,
              inner sep=0pt,rounded corners=1pt] (value) {\small $#1$};}
    \newcommand{\certvariable}[1]{%
        \tikz[baseline=(value.center)]
        \node[draw=black!40,fill=black!8,text=black!65,minimum size=1.45em,
              inner sep=0pt,rounded corners=1pt] (value) {\small $#1$};}
    \newcommand{\certtuple}[3]{#1\kern.02em#2\kern.02em#3}
    \small
    \setlength{\tabcolsep}{.35em}
    \renewcommand{\arraystretch}{1.35}
    \begin{tabular}{c@{\quad}*{11}{c}}
        % & \multicolumn{11}{c}{\textit{variable evolution}}\\[.15em]
        & $a$ & $b$ & $c$ & $d$ & $e$ & $f$ & $g$ & $x$ & $y$ & $z$ & $u$\\
        $s_1$
        & \zerovalue & \fixedvalue{1} & \fixedvalue{2} & \fixedvalue{3}
        & \fixedvalue{4} & \fixedvalue{1} & \fixedvalue{2}
        & \zerovalue & \zerovalue & \zerovalue & \zerovalue\\
        $s_2$
        & \zerovalue
        & \inactivevalue & \inactivevalue & \inactivevalue & \inactivevalue
        & \inactivevalue & \inactivevalue
        & \zerovalue & \fixedvalue{1} & \fixedvalue{4} & \zerovalue\\
        $s_3$
        & \fixedvalue{1}
        & \inactivevalue & \inactivevalue & \inactivevalue & \inactivevalue
        & \inactivevalue & \inactivevalue
        & \fixedvalue{1} & \inactivevalue & \inactivevalue & \fixedvalue{3}\\[.2em]
        \hline
        $\gamma$
        & $1$ & $1$ & $2$ & $3$ & $4$ & $1$ & $2$ & $1$ & $1$ & $4$ & $3$
    \end{tabular}
    \vspace{.8em}

    \renewcommand{\arraystretch}{1.15}
    \begin{tabular}{c@{\quad}*{7}{c@{\hspace{.35em}}}c}
        % & \multicolumn{8}{c}{\textit{constraint evolution}}\\[.15em]
        & $C_1$ & $C_2$ & $C_3$ & $C_4$ & $C_5$ & $C_6$ & $C_7$ & $C_8$\\[-.05em]
        \textit{start}
        & \certtuple{\certvariable{a}}{\certvariable{b}}{\certvariable{e}}
        & \certtuple{\certvariable{a}}{\certvariable{c}}{\certvariable{d}}
        & \certtuple{\certvariable{a}}{\certvariable{d}}{\certvariable{g}}
        & \certtuple{\certvariable{a}}{\certvariable{e}}{\certvariable{f}}
        & \certtuple{\certvariable{b}}{\certvariable{c}}{\certvariable{g}}
        & \certtuple{\certvariable{b}}{\certvariable{d}}{\certvariable{f}}
        & \certtuple{\certvariable{x}}{\certvariable{y}}{\certvariable{z}}
        & \certtuple{\certvariable{a}}{\certvariable{x}}{\certvariable{u}}\\[.3em]
        $s_1$
        & \certtuple{\certvariable{a}}{\certvalue{1}}{\certvalue{4}}
        & \certtuple{\certvariable{a}}{\certvalue{2}}{\certvalue{3}}
        & \certtuple{\certvariable{a}}{\certvalue{3}}{\certvalue{2}}
        & \certtuple{\certvariable{a}}{\certvalue{4}}{\certvalue{1}}
        & \certtuple{\certvalue{1}}{\certvalue{2}}{\certvalue{2}}
        & \certtuple{\certvalue{1}}{\certvalue{3}}{\certvalue{1}}
        & \certtuple{\certvariable{x}}{\certvariable{y}}{\certvariable{z}}
        & \certtuple{\certvariable{a}}{\certvariable{x}}{\certvariable{u}}\\[.3em]
        $s_2$
        & & & & & &
        & \certtuple{\certvariable{x}}{\certvalue{1}}{\certvalue{4}}
        & \certtuple{\certvariable{a}}{\certvariable{x}}{\certvariable{u}}\\[.3em]
        $s_3$
        & & & & & & &
        & \certtuple{\certvalue{1}}{\certvalue{1}}{\certvalue{3}}\\[.2em]
        \hline
        \textit{end}
        & \certtuple{\certvalue{1}}{\certvalue{1}}{\certvalue{4}}
        & \certtuple{\certvalue{1}}{\certvalue{2}}{\certvalue{3}}
        & \certtuple{\certvalue{1}}{\certvalue{3}}{\certvalue{2}}
        & \certtuple{\certvalue{1}}{\certvalue{4}}{\certvalue{1}}
        & \certtuple{\certvalue{1}}{\certvalue{2}}{\certvalue{2}}
        & \certtuple{\certvalue{1}}{\certvalue{3}}{\certvalue{1}}
        & \certtuple{\certvalue{1}}{\certvalue{1}}{\certvalue{4}}
        & \certtuple{\certvalue{1}}{\certvalue{1}}{\certvalue{3}}
    \end{tabular}
    \captionsetup{hypcap=false}
    \caption{An example run of the algorithm.}
    \label{fig:peeling-example}
\end{minipage}
\end{center}
\end{figure}

\paragraph{The algorithm as a reduction.} 
Viewed more generally, the algorithm yields a polynomial-time reduction between the original PCSP and a PCSP obtained by restricting (and thus strengthening) its promise relation. The previous example is the special case in which the restricted promise relation is empty, and thus the target of the reduction is a trivial problem.
The existence of this reduction contributes to the goal of the algebraic approach to PCSPs, which seeks to understand the complexity of all PCSPs through the algebraic characterisation of reductions between them~\cite{BBKO21_Journ_AlgebraicApproachPromiseConstraintSatisfaction,BK22-CombinatorialGapTheoremReductionsPromiseCSPs,DO24_Conf_LocalConsistencyReductionConstraintSatisfactionProblemsa,BHZ27:soda}.
To this end, we show
(\Cref{appendix:implementing-reduction-singleton-algorithms}) that the reduction
can be implemented using \emph{(Constraint) Singleton AIP} ($\CSinglAIP$).
We then \emph{characterise} $\CSinglAIP$ reductions algebraically, along the lines of Dalmau and Opr\v{s}al~\cite{DO24_Conf_LocalConsistencyReductionConstraintSatisfactionProblemsa}
(\Cref{appendix:characterisation-reduction}).

\paragraph{Additive combinatorics.}

Our (hardness) results use tools and ideas from additive combinatorics.
Fix some prime $p\geq 5$. We call a partition of $\Z_p$ into three sets $A_1,A_2,A_3$
\emph{valid} if the only solution to $x+y+z\equiv 0\pmod p$
with $x,y,z$ all in the same part is $(0,0,0)$. Validity is preserved by permuting the parts and by multiplying every
element by the same nonzero scalar.
A first step in the hardness proof is to classify these partitions:
for every prime $p\geq 7$, there is \emph{exactly one} up to these symmetries.

\begin{restatable}{theorem}{partitionsseven}\label{thm:partitions-seven}
    For every prime $p \geq 7$, up to multiplication by a constant and permutation of parts, there exists exactly one valid partition of $\Z_p$, namely the partition that puts $x$ into $A_{1 + \lfloor \frac{3x}{p} \rfloor}$.
\end{restatable}

This fact follows from an application of the Cauchy--Davenport theorem, together
with Vosper's theorem. It does not hold for $p = 5$, as shown by the theorem
below, which can be proved by a simple case analysis. The full proofs of both theorems are in \Cref{appendix:valid-partitions}.

\begin{restatable}{theorem}{partitionsfive}\label{thm:partitions-five}
    Up to multiplication by a constant and permutation of parts, there exist exactly two valid partitions of $\Z_5$, namely 
    $\bigl(\{2,3\},\{4\},\{0,1\}\bigr)$ and $\bigl(\{2,3\},\{1,4\},\{0\}\bigr)$.
\end{restatable}

\section{Preliminaries}

We will only deal with PCSPs with a single relation. Therefore, we will specialise all of our notation and definitions for this case. Furthermore, we will leave the \emph{domains} of relations implicit, for simplicity.\footnote{This does no harm even for the general PCSP case; one could always take the domain of any relational structure to be just the set of all values appearing in any relation.}

For any relation $R$ of arity $r$, let $\dom(R)$ denote the set of values which appear in any tuple of $R$. For $A\subseteq\dom(R)$, write $R|_A=R\cap A^r$ for the restriction of
$R$ to $A$, also called the subrelation induced by $A$. For two relations $R, Q$ of arity $r$, define $R \times Q = \{ ((x_1, y_1), \ldots, (x_r, y_r)) \mid (x_1, \ldots, x_r) \in R, (y_1, \ldots, y_r) \in Q \}$. Note that in general $\dom(R \times Q) \subseteq \dom(R) \times \dom(Q)$, but for \emph{symmetric} $R$ and $Q$ this holds with equality. We define $R^n = R \times \cdots \times R$.

We write $R \to Q$ if there exists $h : \dom(R) \to \dom(Q)$ such that for any $(x_1, \ldots, x_r) \in R$ we have $(h(x_1), \ldots, h(x_r)) \in Q$. We call such a map a \emph{homomorphism}. 
If $R \subseteq Q$ then $R \to Q$. 
Let $[n] = \{1, \ldots, n\}$.

Let $P$ and $Q$ be two relations of the same arity such that $P\to Q$.
An instance $I=(V(I),C(I))$ of $\PCSP(P,Q)$ consists of a set of variables $V(I)=\{x_1,\ldots,x_n\}$ and a set of constraints $C(I)=\{C_1,\ldots,C_m\}$, where each constraint $C_i$ is a tuple of variables $(x_{i,1},\ldots,x_{i,k})$. In this paper, we deal specifically with ternary relations ($k=3$), so every constraint is a triple. For a ternary relation $R$, an \emph{$R$-solution} to $I$ is a map $h:V(I)\to\dom(R)$ such that $(h(x),h(y),h(z))\in R$ for every $(x,y,z)\in C(I)$. For $Z\subseteq V(I)$, the induced subinstance is $I[Z]=(Z,C(I)\cap Z^3)$.

\begin{definition}
    Fix $P \to Q$. \emph{$\dPCSP(P,Q)$} (or simply \emph{$\PCSP(P,Q)$}) asks us, given an instance promised to have either a $P$-solution (YES) or to have no $Q$-solution (NO), to distinguish these two cases.
    \emph{$\sPCSP(P,Q)$} provides us with an instance which has a $P$-solution (that is not part of the input), and asks us to find a $Q$-solution of the instance.
\end{definition}
Note that we require $P \to Q$ for the problem to be well-defined, as otherwise the YES and NO cases overlap. It is well known that $\dPCSP(P,Q)$ reduces to $\sPCSP(P,Q)$~\cite{BBKO21_Journ_AlgebraicApproachPromiseConstraintSatisfaction};
the status of the converse is open.

We next state some simple lemmas that allow us to interpret Abelian embeddings as homomorphisms to certain ``equation relations''.
First, we note that when the relation is finite, one can assume without loss of generality that the target of an Abelian embedding is either the integers or a prime-order cyclic group. This is a direct consequence of the classification of finitely generated Abelian groups. The full proofs of this lemma and the next lemma are in~\Cref{appendix:abelian_embeddings}.

\begin{restatable}{lemma}{simpletargetlemma}\label{lem:abelian-embedding-simple-target}
    Let $A$ be a finite set, $r\geq 1$, and $\emptyset \neq R\subseteq A^r$.
    If $R$ admits an Abelian embedding, then it admits one to some
        $G\in\{\Z\}\cup\{\Z_p\mid p\text{ prime}\}$.
\end{restatable}

For an Abelian group $G$, a set $D$, an element $C \in G$, and a positive integer $c$, we define

\begin{minipage}[t]{0.62\textwidth}
\[
\begin{aligned}[t]
    \Eqn_C(G) &\coloneq \{ (x, y, z) \in G^3 \mid x + y + z = C \} \\
    \Eqn(G) &\coloneq \Eqn_0(G)
        = \{ (x, y, z) \in G^3 \mid x + y + z = 0 \} \\
    \Eqn^*(G) &\coloneq \Eqn(G) \cap \NAE_G.
\end{aligned}
\]
\end{minipage}%
\hfill
\begin{minipage}[t]{0.36\textwidth}
\[
\begin{aligned}
    \NAE_D &\coloneq D^3 \setminus
        \{ (x, x, x) \mid x \in D \} \\
    \NAE_c &\coloneq \NAE_{[c]}
\end{aligned}
\]
\end{minipage}

\medskip
The \emph{primal (Gaifman) graph} of a relation $R$ has vertex set $\dom(R)$,
with two values adjacent if they occur together in a tuple of $R$; we
call $R$ \emph{connected} if this graph is connected.
For symmetric, connected, ternary relations, it is convenient to use
the following homomorphism representation of an Abelian embedding.
\begin{restatable}{lemma}{singlemapreplemma}
    Let $R$ be a ternary symmetric and connected relation, and let $G$ be an Abelian group. Then $R$ admits an Abelian embedding to $G$ if and only if there exist
    $C\in G$ and a nonconstant homomorphism
    $q:R\to\Eqn_C(G)$.
\end{restatable}
Except in \Cref{appendix:abelian_embeddings} (where we prove the two above lemmas), we henceforth use
\emph{Abelian embedding} exclusively in the homomorphism sense above.

\begin{remark}\label{normalisation-abelian-embeddings}
    Let $R$ be a ternary relation, and $q:R\to\Eqn_C(G)$ be a nonconstant homomorphism, where
    $G\in\{\Z\}\cup\{\Z_p\mid p\text{ prime}\}$.
    If $G=\Z$ or
    $G=\Z_p$ with $p\neq3$, then $3q-C:R\to\Eqn(G)$ is a nonconstant homomorphism. Thus we may assume without loss of generality that $C=0$.
    If $G=\Z_3$ and $C=2$, then
    $-q:R\to\Eqn_1(\Z_3)$ is an Abelian embedding, so we may assume
    without loss of generality that $C \in \{0,1\}$.
\end{remark}

\section{Tractability Results}\label{sec:tractability}

Our tractability results will be based on a generic reduction, which targets relational structures with homomorphisms to $\Eqn(G)$. It will also require some extra data which we call a \emph{$c$-rounding}, which we now define.

\begin{definition}[$c$-rounding]
Let $G$ be a finitely generated Abelian group. Write $G^*$ for $G\setminus\{0\}$. A \emph{$c$-rounding} for $G$ is a
map
$\chi:G^*\to[c]$
such that, for every $g_1,g_2,g_3\in G$ satisfying
$g_1+g_2+g_3=0$
and
$(g_1,g_2,g_3)\neq(0,0,0)$,
the nonzero values among
$g_1,g_2,g_3$ do not all receive the same colour under $\chi$. 
\end{definition}

\begin{definition}
    A $c$-rounding $\chi$ is \emph{efficient} if, for some fixed representation of $G$ as a direct sum $\mathbb{Z}^n \oplus tG$ where $tG$ is the torsion subgroup of $G$, the value of $\chi(x_1, \ldots, x_n, g)$ for $x_1, \ldots, x_n \in \mathbb{Z}$ and $g \in tG$ can be computed in polynomial time with respect to $\sum_i \log_2(1 + |x_i|)$, the number of bits needed to represent $x_1, \ldots, x_n$.
\end{definition}

Every $c$-rounding we will give below will be efficient.
Note that an (efficient) $c$-rounding for $G$ is also an (efficient) $c'$-rounding for $G$ for every $c'\geq c$.

Let us now state our goal for this section.

\begin{restatable}{theorem}{reduction}\label{recursive rounding reduction}
    Let $R$ be a finite relation, and $c\geq2$. Suppose there exist a finitely generated Abelian group $G$ as well as
        a homomorphism $q$ from $R$ to $\Eqn(G)$, and
        a $c$-rounding $\chi$ for $G$.
    Then, $\dPCSP(R,\NAE_c)$ reduces to $\dPCSP(R|_{q^{-1}(0)},\NAE_c)$ in polynomial time.
    Furthermore, if $\chi$ is efficient, then
    $\sPCSP(R,\NAE_c)$ reduces to $\sPCSP(R|_{q^{-1}(0)},\NAE_c)$ in polynomial time.
\end{restatable}

\paragraph{Informal description.}
Given an instance $I$ of $\sPCSP(R,\NAE_c)$, let $h$ be a promised
$R$-solution of $I$ (although it is not given to us). Then $q\circ h$ is an $\Eqn(G)$-solution of $I$
which takes values in $q(\dom(R))$.
Starting from the given instance, we repeatedly perform the following
four steps.
\begin{enumerate}
    \item \textit{Solve}\footnote{Here we use that $G$ is finitely generated Abelian:
    decomposing $G$ into its free and torsion parts and encoding the latter
    by integer equations reduces the problem to a linear Diophantine
    system, solvable in polynomial time~\cite[Corollary~5.3b]{Schrijver86}.
    The resulting solution can be represented using a polynomial number of bits.}
    the induced system of equations over $G$, seeking
    a solution $s$ that assigns some variable a nonzero value from
    $q(\dom(R))$. If none exists, then we know that the image of $h$ maps to 0 through $q$. Hence, we can \emph{restrict} the domains of the variables to $q^{-1}(0)$, reducing the problem to $\sPCSP(R|_{q^{-1}(0)},\NAE_c)$ as claimed. So assume we find such a solution $s$.

    \item \textit{Round} each nonzero value $s(x)$ to $\chi(s(x)) \in [c]$; i.e., set variable $x$ to $\chi(s(x))$ once and for all.\footnote{If $\chi$ is efficient, then these values can also be computed in polynomial time, since the torsion part of $s(x)$ is of constant size, and the total number of bits used to represent the nontorsion part of $s(x)$ is polynomial.}

    \item \textit{Peel} off the nonzero support
    $P=\{x\in V(I)\mid s(x)\neq0\}$. By the definition of $c$-rounding, every constraint which intersects $P$ is already satisfied, no matter what values are assigned to the variables outside of $P$. Thus it is safe to remove all constraints intersecting $P$, as well as all variables in $P$. % Every constraint meeting $P$ is

    \item \textit{Recurse} on the subinstance induced by the remaining
    zero set $s^{-1}(0)=V(I)\setminus P$. Since $P\neq\emptyset$, each
    round removes at least one variable.
\end{enumerate}

\begin{remark}
  Since every $R|_{q^{-1}(0)}$-solution is an $R$-solution, $\dPCSP(R|_{q^{-1}(0)},\NAE_c)$ reduces to
    $\dPCSP(R,\NAE_c)$ in polynomial time. Thus, \Cref{recursive rounding reduction} implies a polynomial-time equivalence between $\dPCSP(R|_{q^{-1}(0)},\NAE_c)$ and $\dPCSP(R,\NAE_c)$. The same holds for $\sPCSP$.
\end{remark}

\begin{remark}
For a finite Abelian group \(G\), define the subequation relation
\[
\SubEqn(G)=
\left\{(x,y,z)\in(G^*)^3:
x+y+z=0
\text{ or }
x+y=0
\text{ or }
x+z=0
\text{ or }
y+z=0
\right\}.
\]
Let \(q:R\to\Eqn(G)\), and put $R_0=R|_{q^{-1}(0)}$.
Provided \(R\to \SubEqn(G)\), the proof of
\Cref{recursive rounding reduction}, with each nonzero value \(s(x)\)
retained instead of mapped through a rounding, gives 
$\dPCSP(R,\SubEqn(G))\leq_p\dPCSP(R_0,\SubEqn(G))$
and
$\sPCSP(R,\SubEqn(G))\leq_p\sPCSP(R_0,\SubEqn(G))$.
Indeed, when a constraint is first removed, either its three nonzero
values sum to zero, or exactly two values are nonzero and sum to zero.
Hence it is satisfied in \(\SubEqn(G)\), independently of values assigned
later to the remaining variables.
Moreover, a \(c\)-rounding
\(\chi:G^*\to[c]\) is exactly the same as a homomorphism
$
\SubEqn(G)\to\NAE_c.
$
\end{remark}

\begin{example}\label{rounding examples}
We give a few examples of roundings (and two nonexamples):
    \begin{itemize}
        \item The following is a $2$-rounding for $\mathbb{Z}$:
    \[
        \chi_\infty: \mathbb{Z}^* \to \{1,2\},\qquad\chi_\infty(x) = \begin{cases}
            1, &  x > 0 \\
            2, & x < 0
        \end{cases}
    \]
    \item The following is a $4$-rounding for $\mathbb{Z}_p$ with $p\geq 5$ prime:
    \[
        \chi_p: \mathbb{Z}_p^* \to \{1,2,3,4\},\qquad \chi_p(x) = \begin{cases}
            1, & 0 < x < \frac{p}{3} \\
            2, & \frac{p}{3} < x < \frac{p}{2} \\
            3, & \frac{p}{2} < x < \frac{2p}{3} \\
            4, & \frac{2p}{3} < x < p
        \end{cases}
    \]
    \item More generally, for every prime $p\geq5$ and every $k\geq1$,
    the following is a $4$-rounding for $\Z_{p^k}$. Given
    $x\in\Z_{p^k}^*$, write $x=p^j a$, where $0\leq j<k$ and
    $p\nmid a$, and define
   $\chi_{p^k}(x)=\chi_p(a\bmod p)$.
    To see that this works, write $x$ in base $p$, and for $x + y + z = 0$ consider the least significant nonzero digits in $x, y, z$. At least two must appear at the same position, and must sum up to 0 --- hence correctness follows from the correctness of $\chi_p$.
    
    \item If $\chi$ is a $c$-rounding for $G$, then the following is a $c$-rounding for $G^n$:
    $$
        \chi^{(n)}:\left(G^n\right)^*\to[c],\qquad
        \chi^{(n)}(x_1,\ldots, x_n)=\chi(x_j),
        \quad j=\min\{i\mid x_i\neq0\}.
    $$
    This follows by essentially the same reasoning as the previous example.
    \item More generally, suppose that $\chi_i$ is a $c$-rounding for
    $G_i$, for every $i\in[n]$. Then their direct product has the
    $c$-rounding
    \[
        \chi_\Pi:\left(\prod_{i=1}^nG_i\right)^*\to[c],\qquad
        \chi_\Pi(x_1, \ldots, x_n)=\chi_j(x_j),
        \quad j=\min\{i\mid x_i\neq0\}.
    \]
    
    \item \emph{Nonexamples} involve $\Z_2$ and $\Z_3$ as there are no $c$-roundings for $\Z_2$ and $\Z_3$, for any $c$.
    In $\Z_2$, the equation $1+1+0=0$ has two nonzero entries, both equal to $1$, so they necessarily receive the same colour. In $\Z_3$, the same obstruction is given by $1+1+1=0$. Both equations violate the
    defining property of a rounding.
   \end{itemize}
\end{example}

We defer a formal proof of \Cref{recursive rounding reduction} to \Cref{app:reduction}; the filling in of the informal description is, however, straightforward.

\begin{remark}[Generalising to higher arities]\label{generalisation to higher arities}
For $r\geq3$, define
\[
    \Eqn^{(r)}(G)
      =\left\{(x_1,\ldots,x_r)\in G^r:
        \sum_{i=1}^r x_i=0\right\},
    \qquad
    \NAE_D^{(r)}
      =D^r\setminus\{(a,\ldots,a):a\in D\}, \qquad c_r = \sum_{k=1}^r\varphi(k),
\]
where $\varphi$ is the Euler totient function.
An $r$-ary $c$-rounding of $G$ is a map
$\chi:G\setminus\{0\}\to[c]$ such that, for every nonzero tuple
$(x_1,\ldots,x_r)\in\Eqn^{(r)}(G)$, the nonzero entries receive at
least two different colours. With these replacements, the proof of
\Cref{recursive rounding reduction} applies verbatim in arity $r$.

The existence of a rounding depends on the arity. An $r$-ary
rounding is an $s$-ary rounding for every $s\leq r$, by padding
zero-sum $s$-tuples with zeros, but the converse need not hold.
If $p\leq r$, then $\Z_p$ has no $r$-ary
$c$-rounding for any $c$. If $p>r$ on the other hand, we can generalise the construction of the ternary $4$-rounding to get an $r$-ary $c_r$-rounding for $\Z_p$ as follows: construct the set $
Q_{p,r}=\left\{p\frac ab:1\leq a\leq b\leq r\right\} \subset \mathbb{Q}
$. Note that $|Q_{p,r}| = c_r$ for all $p$. Then, identifying $\Z_p^*$ with $\{1,\dots,p-1\}$, assign to each $x \in \Z_p^*$ the colour indexed by the least element of $Q_{p,r}$ greater than $x$. This is an $r$-ary $c_r$-rounding of $\Z_p$. There are however for many pairs $(p,r)$ roundings which use even fewer colours.

Finally, although \Cref{lem:abelian-embedding-simple-target} generalises immediately, the normalisation of an Abelian embedding described in~\Cref{normalisation-abelian-embeddings} also
depends on the arity: if
$q:R\to\Eqn_C^{(r)}(\Z_p)$, then
$x\mapsto rq(x)-C$ is a nonconstant homomorphism to
$\Eqn^{(r)}(\Z_p)$ provided $p\nmid r$. Primes dividing $r$ must be
treated separately.
\end{remark}
\subsection{Consequences of \texorpdfstring{\Cref{recursive rounding reduction}}{Theorem}}

\begin{corollary}\label{empty zero preimage tractability}
    Under the assumptions of \Cref{recursive rounding reduction} for the $\sPCSP$ reduction, if
    $R|_{q^{-1}(0)}=\emptyset$, then
    $\sPCSP(R,\NAE_c)$ is solvable in polynomial time.
\end{corollary}
\begin{proof}
    By \Cref{recursive rounding reduction}, it suffices to solve
    $\sPCSP(\emptyset,\NAE_c)$. Any promised instance of this problem
    has no constraints, so assigning an arbitrary colour to each
    remaining variable gives an $\NAE_c$-solution.
\end{proof}

Together with the existence of an Abelian embedding to $\Z$ for $\T$, this is enough to prove that $\sPCSP(\T, \NAE_2)$ can be solved in polynomial time, a classical theorem in the literature~\cite{BG21}.
This corollary is also sufficient to prove our tractability claim about detrivialised equations.

\begin{corollary}\label{detrivialised vector equations tractable}
    For any finite Abelian group $G$ without elements of order 2 or 3, $\sPCSP(\Eqn^*(G), \NAE_4)$ is solvable in polynomial time.
\end{corollary}
\begin{proof}
    By the classification theorem for finite Abelian groups, $G = \Z_{p_1^{k_1}}\times \cdots \times \Z_{p_n^{k_n}}$ with $p_i \neq 2, 3$. By \Cref{rounding examples}, $\Z_{p_i^{k_i}}$ has a 4-rounding for $i = 1, \ldots, n$, and the product of groups which admit 4-roundings also admits a 4-rounding. Hence $G$ admits a 4-rounding.
     Moreover, letting $q=\mathrm{id}_G$, $q$ is a homomorphism $ \Eqn^*(G) \to  \Eqn(G)$ with $q^{-1}(0)=\{0\}$, and
    $
        \Eqn^*(G)|_{\{0\}}=\emptyset
    $
    since $(0,0,0)\notin\Eqn^*(G)$. The result now follows
    from \Cref{empty zero preimage tractability}.
\end{proof}

\begin{corollary}\label{recursive abelian embeddings tractable}
    Let $R$ be a finite ternary relation. Suppose that every
    nonempty induced subrelation $S=R|_A$ satisfies at least one of the
    following:
    \begin{enumerate}
        \item there is a nonconstant homomorphism $q$ from $S$ to $\Eqn(\Z)$ or $\Eqn(\Z_p)$ for
        some prime $p>3$;
        \item there is a homomorphism $q:S\to\Eqn_1(\Z_3)$.
    \end{enumerate}
    Then $\sPCSP(R,\NAE_4)$ is solvable in polynomial time.
\end{corollary}
\begin{proof}
    Set $S_0=R$, and consider the current nonempty induced subrelation
    $S_i$. If there is a homomorphism $q_i:S_i\to\Eqn_1(\Z_3)$, solve (by Gaussian elimination) the induced system of equations of the form $x+y+z\equiv1\pmod 3$. It has a solution because the
    instance has an $S_i$-solution. Moreover, since $\Eqn_1(\Z_3)$ contains no constant tuple, any solution is an
    $\NAE_3$-colouring, and hence an $\NAE_4$-colouring.

    Otherwise, there is a nonconstant
    homomorphism $q_i:S_i\to\Eqn(G_i)$, where $G_i=\Z$ or $G_i=\Z_p$ for
    some prime $p\geq5$. In either case $G_i$ has an efficient $4$-rounding (see~\Cref{rounding examples}), so
    \Cref{recursive rounding reduction} reduces the problem to the
    induced subrelation
 $S_{i+1}=S_i|_{q_i^{-1}(0)}$.
    Since $q_i$ is nonconstant, $q_i^{-1}(0)$ is a proper subset of
    $\dom(S_i)$.

    We may repeat this argument because every $S_i$ is an induced
    subrelation of $R$. After at most $|\dom(R)|$ steps we therefore
    encounter the $\Z_3$ case or reach the empty relation. Both have a
    polynomial-time search algorithm, and the reductions lift the resulting
    solution to an $\NAE_4$-solution of the original instance.
\end{proof}
\begin{remark}
    \Cref{recursive abelian embeddings tractable} and its proof extend to
every arity \(r\geq3\): replace \(\Eqn\) by \(\Eqn^{(r)}\),
\(\NAE_4\) by \(\NAE_{c_r}^{(r)}\) (\(c_r\) is as in \Cref{generalisation to higher arities}), the condition \(p>3\) in
item~(1) by \(p>r\), and item~(2) by
\[
    S\to\Eqn^{(r)}_1(\Z_p)
    \qquad\text{for some prime }p\mid r.
\]
\end{remark}

\begin{example}
    Let $R$ be a nonempty symmetric ternary relation, and take two new elements $+,-\not\in\dom(R)$. Define the \emph{suspension} of $R$ as
    $$
    S(R) = R \cup \{(a,+,-) \text{ and all permutations of this tuple} \mid a \in \dom(R)\}.
    $$
    The relation $S(R)$ has an Abelian embedding to every nontrivial Abelian group, given by $+ \mapsto g, - \mapsto -g, a \mapsto 0 \ \forall a \in \dom(R)$ for some nonzero $g\in G$. Moreover, if \(G\) admits a rounding, every nonconstant
\(q:S(R)\to\Eqn(G)\) satisfies \(q^{-1}(0)=\dom(R)\), and hence
\(S(R)|_{q^{-1}(0)}=R\). Thus no choice of embedding can remove more than one suspension layer,
and iterating the construction makes arbitrarily many applications of \Cref{recursive rounding reduction} necessary before the original relation \(R\) is reached.

    For a prime \(p\geq5\), with \(\dom(R)\cap\mathbb Z_p=\emptyset\), define the \emph{$p$-suspension} of $R$ as
    \[
    S_p(R)=R\cup\Eqn^*(\Z_p)\cup
    \left\{
    (a,x,-x)
    \text{ and all its permutations}\mid
    a\in\dom(R),\ x\in\mathbb Z_p^*
    \right\}.
    \]
    Note that $S_p(R)$ has an Abelian embedding to $\Z_p$, given by \(\pi_p(a)=0\) for \(a\in\dom(R)\) and \(\pi_p(x)=x\) for \(x\in\Z_p\). Then \(S_p(R)|_{\pi_p^{-1}(0)}=R\). 
    Moreover, every Abelian embedding \(q:S_p(R)\to\Eqn_C(G)\) has the form
    \(q=g+\varphi\circ\pi_p\), where \(g\in G\) satisfies \(3g=C\), and
    \(\varphi:\Z_p\to G\) is a group homomorphism.

    Hence \(S_p(R)\) embeds into none of \(\Z,\Z_{p'}\) for primes
    \(p'\neq p\), so nested $p$-suspensions can force different prime-order
    targets at successive steps.
\end{example}

\begin{remark}
    One interesting open question is the complexity of $\PCSP(\Eqn^*(\Z_2^n),\NAE_c)$ for $n\geq 2$ and $c$ sufficiently large.\footnote{Note that $\Eqn^*(\Z_p^n)$ is similar to, but not the same as the \emph{Vector valued punctured 3-Lin} relation from~\cite{BKM26_journal_CSPV}. Vector valued punctured 3-Lin corresponds to the restriction of $\Eqn^*(\Z_p^n)$ to $\Z_p^n \setminus \{\vec{0}\}$.} $\Eqn^*(\Z_2^n)$ has an Abelian embedding (to $\Z_2$ for instance), but not the kind that gives tractability in \Cref{recursive abelian embeddings tractable}.
\end{remark}
\section{Hardness of approximating detrivialised equations with 3 colours}\label{sec:hardness}
For the hardness results, we will need the following notions from the algebraic approach to PCSPs~\cite{BBKO21_Journ_AlgebraicApproachPromiseConstraintSatisfaction}, which we specialise to the case of relational structures with one \emph{symmetric} relation.\footnote{Recall that for symmetric $R$, $\dom(R^n) = \dom(R)^n$.}
An $n$-ary \emph{polymorphism} is a homomorphism $f:R^n \to Q$. We let $\Pol^{(n)}(R, Q)$ denote the set of all polymorphisms $R^n \to Q$, and $\Pol(R, Q) = \bigcup_{n \geq 1}^\infty \Pol^{(n)}(R, Q)$.
For this, we need to define the notion of a \emph{minor} of an operation. Suppose $f : A^n \to B$ and $\pi : [n] \to [m]$ are functions. We then define $f^\pi : A^m \to B$ by
\[
    f^\pi(x_1, \ldots, x_m) = f(x_{\pi(1)}, \ldots, x_{\pi(n)}).
\]
\begin{definition}
    A \emph{(function) minion} $\mathcal{M}$ on a pair of finite sets $(A, B)$ is a nonempty subset of $\left\{f: A^n \rightarrow B \mid n \geq 1\right\}$ that is closed under taking minors, meaning that for each minor map $\pi : [n] \to [m]$ and $n$-ary $f \in \M$, we have $f^\pi \in \M$. For fixed $n \geq 1$, let $\mathcal{M}^{(n)}$ denote the set of $n$-ary functions in $\mathcal{M}$.
\end{definition}
Note that for relations $R$ and $Q$, $\Pol(R,Q)$ is closed under taking minors. If there is also a homomorphism from $R$ to $Q$, then $\Pol(R,Q)$ is nonempty and hence forms a minion.
\begin{definition}
    Let $\M$ and $\N$ be minions.
    A collection of maps $\mu = (\mu_n)_{n=1}^\infty$ with $\mu_n:\M^{(n)}\to\N^{(n)}$ is a \emph{minion homomorphism} from $\M$ to $\N$ if it preserves taking minors, meaning that for each minor map $\pi : [n] \to [m]$ and $f \in \M^{(n)}$, we have
    $(\mu_n(f))^\pi = \mu_m(f^\pi)$.
\end{definition}
\begin{theorem}\protect{\cite[Theorem~3.1]{BBKO21_Journ_AlgebraicApproachPromiseConstraintSatisfaction}}\label{minion hom implies reduction}
    Let $R_1,R_2,Q_1,Q_2$ be relations such that $R_1\to Q_1$ and $R_2\to Q_2$. If there exists a minion homomorphism from $\Pol(R_1,Q_1)$ to $\Pol(R_2,Q_2)$, then $\PCSP(R_2,Q_2)$ reduces to $\PCSP(R_1,Q_1)$ in logarithmic space.
\end{theorem}

Our goal is to prove that $\PCSP(\Eqn^*(G), \NAE_3)$ is \NP-hard whenever $G$ has no elements of order 2 or 3, and the problem is well-defined, i.e.~$\Eqn^*(G) \to \NAE_3$. We show that for this it suffices to consider the case $G = \Z_p$ for $p \geq 5$ --- in the subsequent sections we prove hardness for these cases.

\begin{lemma}\label{lem:reduction}
If an Abelian group $G$ admits an element $g$ of order $k$, and $\Eqn^*(G) \to \NAE_3$, then $\PCSP(\Eqn^*(\Z_k), \NAE_3)$ reduces to  $\PCSP(\Eqn^*(G), \NAE_3)$. In particular, if $G$ is a nontrivial finite Abelian group with no elements of order 2 or 3 such that $\Eqn^*(G) \to \NAE_3$, then $\PCSP(\Eqn^*(\Z_p), \NAE_3)$ reduces to $\PCSP(\Eqn^*(G), \NAE_3)$ for some prime $p \geq 5$.
\end{lemma}
\begin{proof}
    Note that $n \mapsto ng$ 
    determines an injective homomorphism $\Z_k \to G$, which induces a homomorphism $\Eqn^*(\Z_k) \to \Eqn^*(G)$ 
    (since injectivity preserves the nonconstancy of tuples). Thus, as every $\Eqn^*(\Z_k)$-solution induces an $\Eqn^*(G)$-solution by composition with this homomorphism, the identity map gives the required reduction. For the second part, a nontrivial finite Abelian group must have some element of prime order, and by assumption it cannot have order 2 or 3.
\end{proof}

The proof of hardness in \Cref{general theorem} then follows by \Cref{lem:reduction} with \Cref{thm 5 hard} and \Cref{thm geq 7 hard}.

\subsection{Hardness for \texorpdfstring{$\bm{p=5}$}{p=5}}
\begin{definition}
    Following \cite{HN04-GraphsHomomorphisms}, a digraph $G$ is
    \emph{balanced} if each of its oriented cycles has zero net length.
    Following \cite{BBB21_Conf_SymmetricPromiseConstraintSatisfactionProblemsBooleanCase,CKKNZ25_Journ_1in3VsNotAllEqualDichotomyBrokenPromise},
    associate with a directed graph $G$, viewed as a binary relation, the
    ternary symmetric relation
    $\widehat{G}
        \coloneq \{(x,x,y),(x,y,x),(y,x,x)\mid (x,y)\in G\}$.
\end{definition}

\begin{theorem}\protect{\cite[Theorem~2]{CKKNZ25_Journ_1in3VsNotAllEqualDichotomyBrokenPromise}}
    Let $G$ be a bipartite directed graph. Then $\PCSP(\widehat{G},\NAE_2)$ is tractable if $G$ is balanced, and it is \NP-hard otherwise.
\end{theorem}

Let
$
    D_4\coloneq\{(x,-2x)\mid x\in\C\}.
$
This is a directed cycle of net length 4, so it is bipartite and not balanced.
\begin{corollary}\label{D4 NAE2 is hard}
    $\PCSP(\widehat{D}_4,\NAE_2)$ is \NP-hard.
\end{corollary}
Moreover, $\widehat{D}_4$ is the restriction of $\Eqn^*(\Z_5)$ to $\C$.

We will construct a minion homomorphism from $\Pol(\Eqn^*(\Z_5),\NAE_3)$ to $\Pol(\widehat{D}_4,\NAE_2)$.
Let $g\in \Pol(\Eqn^*(\Z_5),\NAE_3)$ with $g(0, \ldots, 0)=3$.
\begin{lemma}
    For every $u\in \C^n$, we have
$
|\{g(u),g(-u)\}\cap\{1,2\}| = 1.
$
\end{lemma}
\begin{proof}
Note that for every $n$-ary polymorphism $g$ and every $u \in \C^n$,  $h(t)\coloneq g(tu)$ is a homomorphism from $\Eqn^*(\Z_5)$ to $\NAE_3$. Thus this follows from the classification of homomorphisms in~\Cref{thm:partitions-five}, and the assumption $g(0, \ldots, 0)=3$.
\end{proof}
\begin{definition}
    For $g \in \Pol(\Eqn^*(\Z_5),\NAE_3)$ with $g(0, \ldots, 0)=3$, define \emph{$\tau_g$} by letting $\tau_g(u)\in\{1,2\}$ denote, for each $u \in \C^n$, the unique element in
$\{g(u),g(-u)\}\cap\{1,2\}$.
\end{definition}

\begin{proposition}\label{tau is a polymorphism}
$\tau_g\in\Pol(\widehat{D}_4,\NAE_2)$.

\end{proposition}

\begin{proof}
Suppose for a contradiction that $u,v,w\in \C^n$ satisfy
$
u+v+w=0
$
but 
$\tau_g(u)=
\tau_g(v)=
\tau_g(w)\eqcolon c$.
Then for each $\lambda\in \C$, the triple
$
(\lambda u,\lambda v,\lambda w)
$
is a constraint triple.
Note that for all $u \in \C^n$,
$\tau_g(-u)=\tau_g(u)$,
and $\tau_g(2u)\neq\tau_g(u)$.
Hence every value in the first and fourth rows of the matrix below is in $\{3,c\}$, and every value in the second and third rows is in $\{3,3-c\}$. Consequently, 3 must appear at least once in each row (otherwise the row would be monochromatic, contradicting that $g$ is a polymorphism).
\[
\begin{matrix}
    g(u)&g(v)&g(w)\\
    g(2u)&g(2 v)&g(2 w)\\
    g(3u)&g(3v)&g(3 w)\\
    g(4u)&g(4 v)&g(4 w)
\end{matrix}
\]

By the pigeonhole principle, some column has at least two occurrences of the value $3$, but since also $g(0, \ldots, 0)=3$, this contradicts \Cref{thm:partitions-five}, applied to the homomorphism induced by that column.
\end{proof}

\begin{proposition}\label{tau preserves minors}
    $g\mapsto\tau_g$ preserves minors.
\end{proposition}

\begin{proof}
Since $\{g^\pi(u),g^\pi(-u)\} = \{g(u^\pi),g(-u^\pi)\}$, we have that $\tau_{g^\pi}(u) = \tau_{g}(u^\pi) = (\tau_{g})^\pi(u)$.
\end{proof}
\begin{theorem}\label{minion hom for p=5}
    There exists a minion homomorphism from $\Pol(\Eqn^*(\Z_5),\NAE_3)$ to $\Pol(\widehat{D}_4,\NAE_2)$.
\end{theorem}
\begin{proof}
    Let $\sigma \in S_3$ be the permutation sending $3\mapsto 2\mapsto1\mapsto3$. 
    For $g \in \Pol(\Eqn^*(\Z_5),\NAE_3)$, let $\tilde{g} = \sigma^{g(0, \ldots, 0)}\circ g$. Note that (since $\sigma$ is an automorphism of $\NAE_3$) $\tilde{g} \in \Pol(\Eqn^*(\Z_5),\NAE_3)$. Also, since $\sigma^1(1)=\sigma^2(2)=\sigma^3(3)=3$, we have $\tilde{g}(0, \ldots, 0)=3$.
    Now let $\mu$ be the map on $\Pol(\Eqn^*(\Z_5),\NAE_3)$  given by
    $\mu(g)\coloneq  \tau_{\tilde{g}}$.
    By \Cref{tau is a polymorphism}, $\mu: \Pol(\Eqn^*(\Z_5),\NAE_3)\to\Pol(\widehat{D}_4,\NAE_2)$.
    It remains to show that $\mu$ preserves minors. Using \Cref{tau preserves minors}, it is enough to show $g \mapsto \tilde{g}$ preserves minors.
    \[\tilde{g^\pi} = \sigma^{g^\pi(0, \ldots, 0)}\circ g^\pi = \sigma^{g(0, \ldots, 0)}\circ g^\pi = (\sigma^{g(0, \ldots, 0)}\circ g)^\pi = \tilde{g}^\pi.\] Here the second equality is because $g^\pi(0, \ldots, 0) = g(0, \ldots, 0)$ and the third equality is because automorphisms commute with minors.
\end{proof}
\begin{theorem}\label{thm 5 hard}
    $\PCSP(\Eqn^*(\Z_5),\NAE_3)$ is \NP-hard.
\end{theorem}
\begin{proof}
    \Cref{minion hom for p=5} provides a minion homomorphism which, by~\Cref{minion hom implies reduction}, gives a log-space reduction from $\PCSP(\widehat{D}_4,\NAE_2)$ to $\PCSP(\Eqn^*(\Z_5),\NAE_3)$. By \Cref{D4 NAE2 is hard}, $\PCSP(\widehat{D}_4,\NAE_2)$ is \NP-hard, hence so is $\PCSP(\Eqn^*(\Z_5),\NAE_3)$.
\end{proof}
\subsection{Hardness for \texorpdfstring{$\bm{p\geq 7}$}{p >= 7}}
Now, we will show \NP-hardness of $\PCSP(\Eqn^*(\Z_p),\NAE_3)$ for all primes $p\geq 7$.
First we must recall a few more standard notions from the algebraic approach to PCSPs~\cite{BBKO21_Journ_AlgebraicApproachPromiseConstraintSatisfaction}.
\begin{definition}
Let $\M$ be a minion. The \emph{free structure of $\M$ generated by $\T$} is the relation $F_\M(\T)$ defined as the set of all triples $(f_1,f_2,f_3)$ of functions in $\M^{(2)}$ for which there exists some $g \in \M^{(3)}$ such that the following identities hold for all $x,y$ in the domain of the $f_i$'s:
$$
f_1(x,y)=g(x,y,y),\qquad
f_2(x,y)=g(y,x,y),\qquad
f_3(x,y)=g(y,y,x).
$$
\end{definition}
\begin{theorem}[Special case of Lemma 4.4 in \cite{BBKO21_Journ_AlgebraicApproachPromiseConstraintSatisfaction}]\label{hom from free structure implies minion hom to projections}
    Let $\M$ be a minion. If $F_\M(\T) \to \T$, then there exists a minion homomorphism from $\M$ to $\Pol(\T,\T)$.
\end{theorem}

\begin{definition}
    For relations $R,Q$ and $S \subseteq \dom(R)$, define \emph{$\Pol(R,Q)|_S$} by
    \[\Pol(R,Q)|_S =\bigcup_n\{f|_{S^n} \mid f \in \Pol^{(n)}(R,Q)\}.\]
\end{definition}
\begin{observation}\label{restrictionisminionhom}
    For any $S \subseteq \dom(R)$, restriction defines a minion homomorphism $\Pol(R,Q)\to\Pol(R,Q)|_S$.
\end{observation}
\begin{restatable}{theorem}{essentiallyunary}\label{thm:essentially-unary}
    For $p\geq 7$ prime, every function in $\Pol^{(2)}(\Eqn^*(\Z_p),\NAE_3)|_{\Z_p^*}$ is nonconstant and essentially unary.
\end{restatable}
We prove~\Cref{thm:essentially-unary} in~\Cref{appendix:essentially-unary}, using a proof obtained with GPT-5.6 Sol (cf. AI methodology).
\begin{theorem}\label{thm geq 7 hard}
    For $p\geq 7$ prime, $\PCSP(\Eqn^*(\Z_p),\NAE_3)$ is \NP-hard.
\end{theorem}
\begin{proof}
Let $\mathcal{M}=\Pol(\Eqn^*(\Z_p),\NAE_3)|_{\Z_p^*}$. We show $F_\mathcal{M}(\T) \to \T$. Define the map $\alpha:\mathcal{M}^{(2)}\to \{0,1\}$ as follows:
    \[\alpha(f)= \begin{cases}
    1, &\text{if the first coordinate is essential in $f$,}\\
    0, &\text{if the second coordinate is essential in $f$.}
\end{cases}\]
This map is well defined by~\Cref{thm:essentially-unary}. Now, we must show that $\alpha$ is a homomorphism from $F_\mathcal{M}(\T)$ to $\T$.
$(f_1,f_2,f_3)\in F_\mathcal{M}(\T)$ means $\exists g \in \mathcal{M}^{(3)}$ such that
$$
f_1(x,y)=g(x,y,y),\qquad
f_2(x,y)=g(y,x,y),\qquad
f_3(x,y)=g(y,y,x).
$$
Write $h(x)=g(x,x,x)$. Each of these $f_i$'s is essentially unary by assumption, so for each $i$ either $f_i(x,y)= h(x)$ (and $\alpha(f_i)=1$) or $f_i(x,y)= h(y)$ (and $\alpha(f_i)=0$). We prove that exactly one of the $f_i$'s has $\alpha(f_i)=1$.
    If $\alpha(f_1)=\alpha(f_2)=\alpha(f_3)$, we derive a contradiction by considering where $g$ maps the (rows of the) following tableau:
    \[\begin{matrix}
        -2&1&1\\
        1&-2&1\\
        1&1&-2\\
    \end{matrix}.\]
    If exactly two of $\alpha(f_1),\alpha(f_2),\alpha(f_3)$ equal $1$, then by symmetry we may assume that $\alpha(f_1)=\alpha(f_2)=1, \alpha(f_3)=0$. We derive a contradiction by considering where $g$ maps the following tableau:
    \[\begin{matrix}
        1&-2&-2\\
        -2&1&-2\\
        1&1&4\\
    \end{matrix}.\]
Thus, exactly one binary polymorphism in each tuple of $F_\mathcal{M}(\T)$ gets assigned the value 1 by $\alpha$, which means that $\alpha$ is a homomorphism from $F_\mathcal{M}(\T)$ to $\T$.

Thus, combining \Cref{hom from free structure implies minion hom to
  projections}, \Cref{restrictionisminionhom}, and \Cref{minion hom implies
  reduction} gives a reduction from $\CSP(\T)$ to $\PCSP(\Eqn^*(\Z_p),\NAE_3)$, proving \NP-hardness.
\end{proof}

\appendix

\section{Abelian embeddings}\label[appendix]{appendix:abelian_embeddings}

We prove the two lemmas about Abelian embeddings from the
preliminaries.

\singlemapreplemma*

\begin{proof}
    Let $A = \dom(R)$.
    Suppose first that $f_1,f_2,f_3$ witness an Abelian embedding.  If
    $x$ and $y$ are adjacent in the primal graph, symmetry lets us
    choose $z$ so that both $(x,y,z)$ and $(y,x,z)$ belong to $R$.
    Subtracting the two corresponding equations gives
    \[
        f_1(x)-f_2(x)=f_1(y)-f_2(y).
    \]
    Thus $f_1-f_2$ is constant on every connected component of the
    primal graph, and hence on all of $A=\dom(R)$.  The same argument,
    after permuting coordinates, shows that $f_1-f_3$ is constant.
    Consequently there are constants $a_1,a_2,a_3\in G$ and a map
    $q:A\to G$ such that
    \[
        f_i=q+a_i\qquad(i\in[3]).
    \]
    At least one $f_i$ is nonconstant, so $q$ is nonconstant.  For
    every $(x,y,z)\in R$ we now have
    \[
        q(x)+q(y)+q(z)=-(a_1+a_2+a_3).
    \]
    Taking $C=-(a_1+a_2+a_3)$ proves the second condition.

    Conversely, suppose that $q(x)+q(y)+q(z)=C$ on $R$.  Set
    \[
        f_1=q-C,\qquad f_2=q,\qquad f_3=q.
    \]
    These maps are nonconstant and their coordinate-wise sum is zero
    on every tuple of $R$.
\end{proof}

\simpletargetlemma*

\begin{proof}[Proof of \Cref{lem:abelian-embedding-simple-target}]
    Let $f_1,\ldots,f_r:A\to\Gamma$ witness an Abelian embedding to
    an Abelian group $\Gamma$.

    Fix $a_0\in A$, put
    \[
        g_i(a)=f_i(a)-f_i(a_0),
    \]
    and let $H\leq\Gamma$ be the subgroup generated by all $g_i(a)$, for
    $i\in[r]$ and $a\in A$.  The group $H$ is finitely generated and
    nontrivial: the former follows from finiteness of $A$, and the
    latter from nonconstancy of at least one $f_i$.  Moreover, if
    \[
        C=-\sum_{i=1}^r f_i(a_0),
    \]
    then every $(x_1,\ldots,x_r)\in R$ satisfies
    \[
        \sum_{i=1}^r g_i(x_i)=C.
    \]
    In particular, $C\in H$.

    By the structure theorem for finitely generated Abelian groups,
    there is a nonconstant homomorphism $\varphi:H\to G$, where $G=\Z$ if
    $H$ is infinite, and $G=\Z_p$ for some prime $p$ if $H$ is finite.
    Since the elements $g_i(a)$ generate $H$ and $g_i(a_0)=0$, at least
    one map $\varphi\circ g_i$ is nonconstant. Define
    \[
        f'_1(a)=\varphi(g_1(a))-\varphi(C),
        \qquad
        f'_i(a)=\varphi(g_i(a))\quad(2\leq i\leq r).
    \]
    Again, at least
    one map $f'_i$ is nonconstant. For every tuple $(x_1,\ldots,x_r)\in R$,
    \[
        \sum_{i=1}^r f'_i(x_i)
        =\varphi\left(\sum_{i=1}^r g_i(x_i)-C\right)=0.
    \]
    Hence these maps give an Abelian embedding to $G$.
\end{proof}

\section{Proof of \texorpdfstring{\Cref{recursive rounding reduction}}{main reduction}}\label[appendix]{app:reduction}
\reduction*
\begin{proof}
    
    Let us start with the reduction for $\sPCSP$.
    Fix an efficient $c$-rounding $\chi$ for $G$, and write
$D=q(\dom(R))$ and $       R_0=R|_{q^{-1}(0)}$.
    
    Let $I$ be an instance of $\sPCSP(R,\NAE_c)$, and fix an $R$-solution $h$ of $I$ for the analysis (the algorithm does
    not need access to $h$). We construct a decreasing sequence
    $V_0\supseteq V_1\supseteq\cdots$ of sets of active variables,
    starting with $V_0=V(I)$, together with a partial colouring
    $\gamma:V(I)\setminus V_i\to[c]$. At stage $i$, the active instance
    is the induced instance $I_i=I[V_i]$. Notice that $h|_{V_i}$ is an
    $R$-solution of $I_i$.

    For every $x\in V_i$ and every $a\in D\setminus\{0\}$, we test
    whether $I_i$ has an $\Eqn(G)$-solution $s_i$ satisfying
    $s_i(x)=a$. If such a solution is found, let
   $P_i=\{x\in V_i\mid s_i(x)\neq0\}$
    be its nonzero support, and set
    $V_{i+1}=V_i\setminus P_i=s_i^{-1}(0)$. For each $x\in P_i$,
    extend the partial colouring by setting
  $\gamma(x)=\chi(s_i(x))$.

    We claim that every constraint removed at this stage is already
    satisfied by $\gamma$. Indeed, let $(x,y,z)$ be a constraint of
    $I_i$ which is not a constraint of $I_{i+1}$. Since $s_i$ is an
    $\Eqn(G)$-solution, $s_i(x)+s_i(y)+s_i(z)=0$.

    Moreover, at least one of $x,y,z$ belongs to $P_i$, so these three
    values are not all zero. The defining property of $\chi$ therefore
    implies that the nonzero values among them receive at least two
    different colours. All variables carrying these nonzero values
    belong to $P_i$, and hence their colours have just been fixed by
    $\gamma$. Thus $(x,y,z)$ is nonmonochromatic, independently of how
    the remaining variables are coloured later.

    Since $P_i$ is nonempty, every successful stage strictly decreases
    the number of active variables. The process therefore reaches,
    after at most $|V(I)|$ stages, an active set $V_t$ for which none of
    the tests above succeeds. We output the residual instance
   $I'=I[V_t]$.
    This is a valid instance of $\sPCSP(R_0,\NAE_c)$. To see this,
    consider the $R$-solution $h|_{V_t}$ of $I'$. The composition
    $q\circ h|_{V_t}$ is an $\Eqn(G)$-solution of $I'$ taking values in
    $D$. If $q(h(x))\neq0$ for some $x\in V_t$, then the test for the
    pair $(x,q(h(x)))$ would have succeeded. Consequently
    $q(h(x))=0$ for every $x\in V_t$, and hence $h|_{V_t}$ is an
    $R_0$-solution of $I'$.

    Given any $\NAE_c$-solution $\gamma'$ of $I'$, extend the partial
    colouring $\gamma$ by setting $\gamma|_{V_t}=\gamma'$. Every constraint
    of $I'$ is satisfied by $\gamma'$, while every other constraint was
    already made nonmonochromatic at the stage when it was removed.
    Therefore $\gamma$ is an $\NAE_c$-solution of $I$.

    Finally, all steps run in polynomial time. Since $G$ is a fixed
    finitely generated Abelian group, systems of linear equations over
    $G$ can be solved in polynomial time. At each of at most $|V(I)|$ stages, we solve one such system
    for every pair $(x,a)\in V_i\times(D\setminus\{0\})$. The set $D$
    is fixed and finite, and $\chi$ is polynomial-time computable by
    assumption. The construction of $I'$ and the lifting of
    $\gamma'$ are therefore both polynomial-time.

    For the decision reduction, the residual instance \(I'\) is computed using only the solutions to the group equations and their zero supports; the algorithm does not evaluate \(\chi\). If \(I\) has an \(R\)-solution, the argument above shows that \(I'\) has an \(R_0\)-solution. Conversely, if \(I'\) has an \(\NAE_c\)-solution, then this solution extends existentially, using the fixed rounding \(\chi\), to an \(\NAE_c\)-solution of \(I\). Thus \no{} instances are preserved, even when \(\chi\) is not efficiently computable.
\end{proof}
\section{Comparison with known algorithms}\label[appendix]{appendix:comparison-known-algos}

\subsection{Failure of k-consistency and BLP+AIP for approximating detrivialised equations}\label[appendix]{appendix:failure-k-cons-BLPAIP}
An $m$-ary
polymorphism $f$ is a \emph{weak near-unanimity} (WNU) polymorphism if
\[
 f(y,x,\ldots,x)=f(x,y,x,\ldots,x)=\cdots=f(x,\ldots,x,y)
\]
for all $x,y$ in the left-hand domain. Atserias and Dalmau proved that
the following condition is necessary even for sublinear width.

\begin{theorem}[{\cite[Corollary~2]{AD22_Conf_PromiseConstraintSatisfactionWidth}}]\label{thm:bounded-width-implies-wnu}
    Let $(R,Q)$ be a PCSP template. If $\PCSP(R,Q)$ is solvable by a
    consistency algorithm of bounded or sublinear width, then
    $\Pol(R,Q)$ contains an $m$-ary WNU polymorphism for every
    $m\geq3$.
\end{theorem}

\begin{proposition}\label{prop:no-bounded-width-detrivialised-equations}
    For every prime $p\neq3$ and every $c$ for which the template is
    defined, $\PCSP(\Eqn^*(\Z_p),\NAE_c)$ is not solvable by
    $k$-consistency for any fixed $k$.
\end{proposition}
\begin{proof}
    It suffices to rule out a ternary WNU polymorphism. Suppose that
    $f\in\Pol^{(3)}(\Eqn^*(\Z_p),\NAE_c)$ were WNU. Since $p\neq3$,
    we have $1\neq-2$, and each column of
    \[
        \begin{pmatrix}
        -2&1&1\\
        1&-2&1\\
        1&1&-2
        \end{pmatrix}
    \]
    is a permutation of the nonconstant zero-sum tuple $(-2,1,1)$ and
    hence belongs to $\Eqn^*(\Z_p)$. The polymorphism property says that
    the tuple obtained by applying $f$ to the rows belongs to $\NAE_c$.
    But the WNU identities make its three entries
    $f(-2,1,1),f(1,-2,1),f(1,1,-2)$ equal, a contradiction. Thus no
    ternary WNU polymorphism exists, so
    \Cref{thm:bounded-width-implies-wnu} rules out bounded-width,
    equivalently fixed $k$-consistency, algorithms.
\end{proof}

We say that a polymorphism of arity $2k + 1$ is \emph{block symmetric} if its outputs are invariant under permutations of its first $k$ inputs and of its last $k + 1$ inputs. We recall the following classical result.

\begin{theorem}[\cite{BGWZ20_Journ_PowerCombinedBasicLPAffineRelaxationPromiseCSPs}]\label{thm:blpaip-characterisation}
    $\PCSP(R,Q)$ is solved by BLP+AIP if and only if
    $\Pol(R,Q)$ contains a block-symmetric polymorphism of arity
    $2k+1$ for every $k\geq1$.
\end{theorem}

\begin{proposition}\label{prop:no-blpaip-detrivialised-equations}
    For every prime $p\geq5$,
    $\PCSP(\Eqn^*(\Z_p),\NAE_c)$ is not solved by BLP+AIP, for every
    $c$ for which the template is defined.
\end{proposition}
\begin{proof}
    Suppose that $f$ is a block-symmetric polymorphism for $k=5$, so
    that its two blocks have sizes $5$ and $6$. Consider the following
    three inputs to $f$, where the vertical line separates the two
    blocks:
    \[
    \left(\begin{array}{ccccc|cccccc}
        \frac{1}{2}&-1&0&-\frac{1}{2}&1
        &1&1&-2&1&1&-2\\
        \frac{1}{2}&0&-\frac{1}{2}&1&-1
        &1&-2&1&1&-2&1\\
        -1&1&\frac{1}{2}&-\frac{1}{2}&0
        &-2&1&1&-2&1&1
    \end{array}\right).
    \]
    Every column is a nonconstant zero-sum tuple, hence belongs to
    $\Eqn^*(\Z_p)$. Within each block, the three rows are permutations
    of one another. Thus block symmetry makes their three images under
    $f$ equal, contradicting preservation of $\NAE_c$.

    Consequently there is no block-symmetric polymorphism for $k=5$,
    and the result follows from
    \Cref{thm:blpaip-characterisation}.
\end{proof}

\subsection{Implementing the reduction with SinglAIP or CSinglAIP}
\label[appendix]{appendix:implementing-reduction-singleton-algorithms}

We begin by recalling AIP and the two singleton variants of AIP studied
recently by
Zhuk~\cite{Zhuk26:focs}.
View a finite-domain CSP instance $I$ as having a current domain $D_x$
for every variable $x$ and a current relation $R_C$ for every
constraint $C \in C(I)$.  For $a\in D_x$, let $I[x=a]$ denote the instance
obtained by replacing $D_x$ with $\{a\}$ and deleting all incompatible
tuples from every constraint containing $x$.  For a constraint
$C=R_C(x_1,\ldots,x_r)$ and a tuple $\bar a\in R_C$, let
$I[C=\bar a]$ denote the instance obtained by replacing the relation
of this particular constraint with $\{\bar a\}$.

\begin{definition}[Affine Integer Programming]
    The \emph{AIP relaxation} of $I$ has integer variables
    $\lambda_x(a)$ for $x\in V(I)$ and $a\in D_x$, and
    $\lambda_C(\bar a)$ for every constraint $C$ and
    $\bar a\in R_C$.  They satisfy
    \[
        \sum_{a\in D_x}\lambda_x(a)=1,
        \qquad
        \sum_{\bar a\in R_C}\lambda_C(\bar a)=1,
    \]
    together with the marginal equations
    \[
        \lambda_{x_i}(a)
        =\sum_{\substack{\bar b\in R_C\\b_i=a}}
        \lambda_C(\bar b)
    \]
    for every $C=R_C(x_1,\ldots,x_r)$, $i\in[r]$, and
    $a\in D_{x_i}$.  There are no nonnegativity constraints.  AIP
    accepts $I$ if this system has an integer solution, and rejects it
    otherwise.
\end{definition}

\begin{definition}[SinglAIP and CSinglAIP]
    Both algorithms repeatedly prune the instance until reaching a
    fixed point.
    \begin{itemize}
        \item \emph{SinglAIP} runs AIP on $I[x=a]$ for every variable
        $x$ and every $a\in D_x$.  If this affine relaxation is
        infeasible, it deletes $a$ from $D_x$ and deletes all tuples
        containing this value from every constraint incident with
        $x$.  It rejects if some domain becomes empty.

        \item \emph{CSinglAIP} runs AIP on $I[C=\bar a]$ for every
        constraint $C$ and every $\bar a\in R_C$.  If this affine
        relaxation is infeasible, it deletes $\bar a$ from the
        relation $R_C$ of this constraint only.  It rejects if some
        constraint relation becomes empty.
    \end{itemize}
    If no further deletion is possible, the corresponding algorithm
    accepts and returns the resulting reduced instance.
\end{definition}

Both algorithms are sound: a value, respectively a constraint tuple,
used by a genuine solution can never be deleted.

The reduction in \Cref{recursive rounding reduction} repeatedly solves
systems of equations and rounds one solution at a time.  We next show
that the same reduction can be expressed as one run of either
Singleton AIP (SinglAIP) or Constraint Singleton AIP (CSinglAIP).

\begin{proof}[Proof of \Cref{recursive rounding reduction} using CSinglAIP]
    Fix an efficient $c$-rounding $\chi$ for $G$, and write $R_0=R|_{q^{-1}(0)}$.
    Let $I$ be an instance of $\sPCSP(R,\NAE_c)$, and fix an
    $R$-solution $h$ of $I$ for the analysis (the algorithm does not
    need access to $h$). Colour variables that occur in no constraint arbitrarily, so we may assume that
    every variable occurs in a constraint. Run CSinglAIP on $I$ to a
    fixed point, obtaining an instance $I^\star$ with nonempty
    surviving constraint relations $R_C^\star$.

    Define the set of surviving constraint-tuple pairs
    \[
        \Omega=\{(C,\bar a)\mid C\in C(I),\ \bar a\in R_C^\star\}.
    \]
    For each $(C,\bar a)\in\Omega$, choose an AIP solution
    $\lambda^{[C=\bar a]}$ of $I^\star[C=\bar a]$, and define the
    induced $\Eqn(G)$-solution
    \[
        s^{[C=\bar a]}(y)
        =\sum_{b\in D_y}\lambda^{[C=\bar a]}_y(b)q(b).
    \]
    If $C=(x_1,x_2,x_3)$ and
    $\bar a=(a_1,a_2,a_3)$, then for $i \in [3]$,
    \[
        s^{[C=\bar a]}(x_i)=q(a_i).
    \]
    Moreover, for every constraint $B=(y_1,y_2,y_3)$, the AIP
    marginal equations give
    \[
        \sum_{i=1}^3s^{[C=\bar a]}(y_i)
        =
        \sum_{\bar b\in R_B^\star}
        \lambda^{[C=\bar a]}_B(\bar b)
        \sum_{i=1}^3q(b_i)
        =0.
    \]
    Thus each $s^{[C=\bar a]}$ is indeed an $\Eqn(G)$-solution of
    $I$.

    Fix an ordering of $\Omega$ and combine these solutions into the
    solution vector
    \[
        \mathbf{s}(y)
        =\bigl(s^{[C=\bar a]}(y)\bigr)_{(C,\bar a)\in\Omega}
        \in G^{|\Omega|}.
    \]
    This is an $\Eqn(G^{|\Omega|})$-solution of $I$. Let
    \[
        P=\{y\in V(I)\mid \mathbf{s}(y)\neq0\},
        \qquad
        Z=V(I)\setminus P=\mathbf{s}^{-1}(0),
    \]
    and let $I'=I[Z]$. For each $y\in P$, define a partial colouring
    by
    \[
        \gamma(y)=\chi^{(|\Omega|)}(\mathbf{s}(y)).
    \]

    Since $\mathbf{s}$ is an $\Eqn(G^{|\Omega|})$-solution and
    $\chi^{(|\Omega|)}$ is a $c$-rounding for $G^{|\Omega|}$, every
    constraint not contained in $I'$ contains two variables in $P$
    that receive different colours under $\gamma$.

    It remains to check that $I'$ is a valid instance of
    $\sPCSP(R_0,\NAE_c)$. Let $y\in Z$, choose a constraint
    $C=(x_1,x_2,x_3)$ containing $y$, and write $y=x_i$. Since
    CSinglAIP never deletes the tuple
    \[
        \bar h_C=(h(x_1),h(x_2),h(x_3)),
    \]
    the pair $(C,\bar h_C)$ belongs to $\Omega$. The corresponding
    coordinate of $\mathbf{s}(y)$ is
    \[
        s^{[C=\bar h_C]}(y)=q(h(y)).
    \]
    Since $\mathbf{s}(y)=0$, it follows that $q(h(y))=0$. Therefore
    $h|_Z$ is an $R_0$-solution of $I'$.

    Given any $\NAE_c$-solution $\gamma'$ of $I'$, extend the partial
    colouring $\gamma$ by setting $\gamma|_Z=\gamma'$. Every
    constraint of $I'$ is satisfied by $\gamma'$, while every other
    constraint was already made nonmonochromatic. Therefore $\gamma$
    is an $\NAE_c$-solution of $I$; assigning arbitrary colours to the
    isolated variables set aside at the beginning completes the
    solution of the original instance.

    Finally, the construction runs in polynomial time. Indeed,
    \[
        |\Omega|\leq |C(I)|\,|R|.
    \]
    CSinglAIP and the extraction of the AIP solutions
    $\lambda^{[C=\bar a]}$ take polynomial time, and the product
    rounding $\chi^{(|\Omega|)}$ is polynomial-time computable.

    For the decision reduction, if CSinglAIP rejects, we output a fixed instance with no \(\NAE_c\)-solution; this is safe because CSinglAIP cannot reject an \(R\)-solvable instance. Otherwise, the residual instance \(I'=I[Z]\) is computed using only the induced group-valued solutions and their common zero set, without evaluating \(\chi^{(|\Omega|)}\). If \(I\) has an \(R\)-solution, the argument above shows that \(I'\) has an \(R_0\)-solution. Conversely, any \(\NAE_c\)-solution of \(I'\) extends existentially, using the fixed product rounding \(\chi^{(|\Omega|)}\), to an \(\NAE_c\)-solution of \(I\). Thus \no{} instances are preserved even when \(\chi\) is not efficiently computable.
\end{proof}
\begin{remark}
    The same reduction can also be implemented using SinglAIP. Indeed, after running SinglAIP to a fixed point, index the pinned AIP solutions by the surviving variable-value pairs \((x,a)\). The corresponding coordinate solution satisfies \(s^{[x=a]}(x)=q(a)\). Hence, if the combined solution vector vanishes at \(x\), then every surviving value \(a\in D_x\) belongs to \(q^{-1}(0)\). The rounding and lifting arguments are otherwise identical. This is also consistent with the fact that every finite PCSP template solved by CSinglAIP is solved by SinglAIP.
\end{remark}
\begin{remark}
    Both SinglAIP and CSinglAIP are idempotent: their outputs are fixed
    points, so rerunning the same algorithm makes no further
    deletions. Consequently, a single run of either singleton algorithm can
    provide several successive applications of
    \Cref{recursive rounding reduction}.
\end{remark}

\subsection{A minion characterisation of CSinglAIP reductions}\label[appendix]{appendix:characterisation-reduction}
We now show that the reduction of \Cref{recursive rounding reduction} can also be performed by the broader class of $\CSinglAIP$ reductions.
Following Dalmau and Opr\v{s}al, we characterise when there exists a $\CSinglAIP$ reduction between PCSPs by a minion condition. 
The characterisation combines Zhuk's~\cite{Zhuk26:focs} skeletal-matrix
description of CSinglAIP, building on~\cite{cz23sicomp:clap}, with the minion formulation of arc-consistency
reductions from~\cite{DO24_Conf_LocalConsistencyReductionConstraintSatisfactionProblemsa}.

We pass from function minions to the more general notion of an abstract minion. We also allow arities to be indexed by arbitrary
nonempty finite sets, rather than only by sets of the form $[n]$.

\begin{definition}[Abstract minion]
    An \emph{(abstract) minion} $\N$ consists of a set $\N^{(X)}$ for
    every nonempty finite set $X$ and, for every map $\pi:X\to Y$, an
    operation
    \[
        \N^{(X)}\to\N^{(Y)},\qquad f\mapsto f^\pi,
    \]
    such that
    \[
        f^{\mathrm{id}_X}=f
        \qquad\text{and}\qquad
        (f^\pi)^\rho=f^{\rho\circ\pi}
    \]
    for all maps $\pi:X\to Y$ and $\rho:Y\to Z$.  The elements of
    $\N^{(X)}$ are said to have arity $X$, and the operations
    $f\mapsto f^\pi$ are called \emph{minors}.

    A \emph{minion homomorphism} $\xi:\N\to\M$ is a collection of
    maps $\xi_X:\N^{(X)}\to\M^{(X)}$ that preserve minors: for every
    $f\in\N^{(X)}$ and $\pi:X\to Y$,
    \[
        \xi_Y(f^\pi)=\xi_X(f)^\pi.
    \]
\end{definition}

A function minion is viewed in this form by taking $\N^{(X)}$ to
consist of functions whose coordinates are indexed by $X$; the minor
corresponding to $\pi:X\to Y$ identifies, permutes, or introduces
dummy coordinates according to $\pi$.

We first define the minion of affine certificates arising from
CSinglAIP.  This is a special case of the skeletal-matrix descriptions
of singleton algorithms from~\cite{Zhuk26:focs}.

\begin{definition}[Skeletal affine minion]
    For a nonempty finite set $X$, let
    $\mathcal{M}_{\text{CSinglAIP}}^{(X)}$ consist of the integer matrices
    \[
        M:X\times J\to\Z,
    \]
    where $J$ is an arbitrary nonempty finite set of columns, subject
    to the following two conditions:
    \begin{enumerate}
        \item every column has sum one, so
        \[
            \sum_{x\in X}M(x,j)=1
            \qquad(j\in J);
        \]
        \item every row occurring nontrivially in $M$ has a unit
        column.  More precisely, if
        \[
            \operatorname{supp}M
            =\{x\in X:M(x,j)\neq0\text{ for some }j\in J\},
        \]
        then, for every $x\in\operatorname{supp}M$, some column of
        $M$ is the unit vector $e_x$.
    \end{enumerate}
    For a map $\pi:X\to Y$, define
    \[
        M^\pi(y,j)=\sum_{x\in\pi^{-1}(y)}M(x,j).
    \]
\end{definition}

The second condition ensures that support cannot disappear through
cancellation when rows are identified: a unit column $e_x$ becomes
$e_{\pi(x)}$.  Consequently,
\[
    \operatorname{supp}(M^\pi)=\pi(\operatorname{supp}M).
\]
The column-sum condition is also preserved by minors, and the unit
columns remain unit columns after rows are identified.  Hence
$\mathcal{M}_{\text{CSinglAIP}}$ is indeed an abstract minion, and the support map describes a minion homomorphism from $\mathcal{M}_{\text{CSinglAIP}}$ to the nonempty-powerset minion
\[
    \mathcal{P}_+^{(X)}=\{S\subseteq X:S\neq\emptyset\},
    \qquad S^\pi=\pi(S).
\]
Dalmau and Opr\v{s}al's construction is
\[
    \omega(\N)^{(X)}
    =\{(S,f):S \in \mathcal{P}_+^{(X)},\ f\in\N^{(S)}\},
\quad
    (S,f)^\pi=\bigl(\pi(S),f^{\pi|_S}\bigr).
\]
\begin{definition}
    For an abstract minion $\N$, define
    \[
        \omega_{\text{CSinglAIP}}(\N)^{(X)}
        =\{(M,f):M\in\mathcal{M}_{\text{CSinglAIP}}^{(X)},\
          f\in\N^{(\operatorname{supp}M)}\},
\quad
        (M,f)^\pi
        =\bigl(M^\pi,f^{\pi|_{\operatorname{supp}M}}\bigr).
    \]
\end{definition}

Mirroring Dalmau and Opr\v{s}al's treatment of arc consistency, we
describe the construction in the language of Label Cover and minions.
A finite \emph{Label Cover instance} $\mathcal L$ consists of a finite
directed multigraph, a nonempty finite domain $X_v$ for every vertex
$v$, and, for every directed edge
$e=(v\xrightarrow{\pi_e}w)$, a projection map
$\pi_e:X_v\to X_w$.  We call $v$ the source and $w$
the target of $e$ and view the edge as the binary constraint with
source coordinate first and relation
\[
    R_e=\{(a,\pi_e(a)):a\in X_v\}\subseteq X_v\times X_w.
\]

Let $\N$ be an abstract minion.  An \emph{$\N$-solution} of
$\mathcal L$ assigns an element $f_v\in\N^{(X_v)}$ to every vertex
$v$ such that
\[
    f_w=f_v^{\pi_e}
\]
for every edge $e=(v\xrightarrow{\pi_e}w)$.  We write
$\mathcal L\to\N$ if such an $\N$-solution exists.

For a CSP instance $I$, its \emph{associated LC instance}, denoted
$\operatorname{LC}(I)$, has a variable of domain $D_x$ for every
variable $x$ of $I$, a variable of domain $R_C$ for every constraint
$C$, and the directed projection constraint
$C\xrightarrow{\pi_i}x$, where $\pi_i:R_C\to D_x$ is the $i$th
coordinate projection, whenever $x$ occurs in position $i$ of $C$.

For any finite CSP instance $J$, let $\kCSinglAIP(J)$ denote the
instance obtained by running CSinglAIP to a fixed point and then
restricting each domain to the values occurring in the surviving
constraint tuples; the domain of an isolated variable is left
unchanged. If some constraint relation becomes empty, regard this
instance as unsatisfiable. The following lemma is the analogue of \cite[Lemma~C.1]{DO24_Conf_LocalConsistencyReductionConstraintSatisfactionProblemsa}.

\begin{lemma}
    \label{lem:adjoints}
    For every finite Label Cover instance $\mathcal L$ and every
    abstract minion $\N$,
    \[
        \kCSinglAIP(\mathcal L)\longrightarrow\N
        \quad\Longleftrightarrow\quad
        \mathcal L\longrightarrow\omega_{\text{CSinglAIP}}(\N).
    \]
\end{lemma}
\begin{proof}
    Write $F_v\subseteq X_v$ for the domain of $v$ in
    $\kCSinglAIP(\mathcal L)$, provided CSinglAIP does not reject.
    Note that for every $a\in F_v$, the AIP relaxation of
    $\kCSinglAIP(\mathcal L)[v=a]$ is feasible.  This also implies that
    for every edge $e=(v\xrightarrow{\pi_e}w)$,
    $
        \pi_e(F_v)=F_w.
    $

    Suppose first that $h:\kCSinglAIP(\mathcal L)\to\N$ is an $\N$-solution sending $v$ to $h_v$.
    We define a map $g:\mathcal L\longrightarrow\omega_{\text{CSinglAIP}}(\N)$ by 
    \[
    g(v)=(M_v,h_v),
    \]
    where $M_v\in \mathcal{M}_\CSinglAIP^{(X_v)}$ is constructed as follows. Let $\Gamma = \{(u,a) \mid u \in V(\mathcal L), a \in F_u\}$. By assumption, we can choose for each $(u,a)\in \Gamma$ an AIP solution to the LC instance $\kCSinglAIP(\mathcal L)[u=a]$, say $\lambda^{u,a}$. We write $\lambda^{u,a}_v$ for its marginal distribution on $v$, a $\Z$-valued vector indexed by $F_v$. We construct $M_v\in \mathcal{M}_\CSinglAIP^{(X_v)}$ as the $X_v \times \Gamma$ matrix whose column indexed by $(u,a)$ is $\lambda^{u,a}_v$ padded with zeros in the entries corresponding to $X_v\setminus F_v$.

    It is easy to check that $M_v \in \mathcal{M}_\CSinglAIP^{(X_v)}$, $\operatorname{supp}{M_v}=F_v$ and $h_v \in \mathcal{N}^{(F_v)}$, and that $g$ respects LC constraints/minors.

    For the reverse implication, let
    $g:\mathcal L\to\omega_{\text{CSinglAIP}}(\N)$, and write
    \[
        g(v)=(M_v,g_v),
        \qquad S_v=\operatorname{supp}M_v.
    \]
    Because $g$ is an $\omega_{\text{CSinglAIP}}(\N)$-solution,
    \[
        M_w=M_v^{\pi_e},
        \qquad S_w=\pi_e(S_v),
        \qquad
        g_w=g_v^{\pi_e|_{S_v}}
    \]
    for every edge $e=(v\xrightarrow{\pi_e}w)$. Note that since taking the minor of a matrix does not change the column index set, all matrices in a connected component have the same column index set.

    We claim that, for every edge
    $e=(v\xrightarrow{\pi_e}w)$ and every $a\in S_v$, the ordered
    tuple $(a,\pi_e(a))$ is never removed by CSinglAIP.  

    Suppose such a tuple is deleted at some point, say the tuple
    $(a,\pi_e(a))$ on $e=(v\xrightarrow{\pi_e}w)$ is the first such to be deleted.  Fix a
    column $j$ of $M_v$ equal to the unit vector $e_a$.   If $x$ is in the same component as $v$, set $j_x = j$. For every other component $K$, choose some $j_K$ in its column index set arbitrarily, and set $j_x=j_K$ for all $x$ in $K$.
    Now define
    \[
        \lambda_x(b)\coloneq M_x(b,j_x),\qquad
        \lambda_d(b,\pi_d(b))\coloneq M_x(b,j_x)
        \quad\bigl(d=(x\xrightarrow{\pi_d}y)\bigr).
    \]
    $\lambda$ is a valid solution to the AIP relaxation of the remaining LC instance: a nonzero coefficient requires $b\in S_x$, and by
    the choice of the first deletion the tuple
    $(b,\pi_d(b))$ is still present.  Every column has sum one, while
    $M_y=M_x^{\pi_d}$ gives the marginal equation at the target of
    $d$; the marginal equation at its source is immediate. Moreover,
    on the distinguished edge $e$, the choice of the unit column gives
    \[
        \lambda_e(a,\pi_e(a))=1
        \quad\text{and}\quad
        \lambda_e(b,\pi_e(b))=0\quad(b\neq a).
    \]
    Thus the AIP relaxation with
    $(a,\pi_e(a))$ forced on $e$ is feasible, contradicting the deletion.
    This proves the claim.

    Since every $S_v$ is nonempty, the claim shows that every edge
    retains at least one tuple; hence CSinglAIP does not reject.
    Therefore the sets $F_v$ are defined.  Moreover,
    $S_v\subseteq F_v$ for every $v$: this follows directly from the
    claim when $v$ is the source of an edge, from
    $S_v=\pi_e(S_u)\subseteq\pi_e(F_u)=F_v$ when $v$ is only a target,
    and from $F_v=X_v$ when $v$ is isolated.

    Define $h_v\in\N^{(F_v)}$ to be the minor of
    $g_v\in\N^{(S_v)}$ obtained by adding dummy coordinates indexed by
    $F_v\setminus S_v$.  Using $S_w=\pi_e(S_v)$ and
    $g_w=g_v^{\pi_e|_{S_v}}$, it follows that
    $h:\kCSinglAIP(\mathcal L)\to\N$ is an $\N$-solution, as required.
\end{proof}

For a finite Label Cover instance $\mathcal L$ and a relation $S$, let
$\pi_S(\mathcal L)$ denote the \emph{universal $S$-gadget} (\cite[Definition~A.13]{DO24_Conf_LocalConsistencyReductionConstraintSatisfactionProblemsa}, presented as the indicator instance in \cite{BBKO21_Journ_AlgebraicApproachPromiseConstraintSatisfaction}): take a
copy of $S^{X_v}$ for every vertex $v$, and for every edge
$e=(v\xrightarrow{\pi_e}w)$ identify each $b\in S^{X_w}$ with
$b\circ\pi_e\in S^{X_v}$.  By~\cite[Lemma~3.16]{BBKO21_Journ_AlgebraicApproachPromiseConstraintSatisfaction},
\[
    \pi_S(\mathcal L)\to T
    \quad\Longleftrightarrow\quad
    \mathcal L\to\Pol(S,T).
\]
\begin{definition}
Let $J_\circ$ be the instance with one variable $x$ and the
single constraint $(x,x,x)$. Define
\[
    \Phi_S(I)=
    \begin{cases}
        J_\circ,
        &\text{if $\CSinglAIP$ rejects on $\operatorname{LC}(I)$},\\
        \pi_S\bigl(\kCSinglAIP(\operatorname{LC}(I))\bigr),
        &\text{otherwise}.
    \end{cases}
\]
We write
\[
    \PCSP(R,Q)\leq_{\CSinglAIP}\PCSP(S,T)
\]
if $\Phi_S$ is a reduction from $\PCSP(R,Q)$ to $\PCSP(S,T)$.
\end{definition}

\begin{definition}[Free structure~\cite{BBKO21_Journ_AlgebraicApproachPromiseConstraintSatisfaction}]
    Let $R\subseteq A^k$ be a nonempty relation on a finite set $A$,
    and let $\M$ be an abstract minion.  The \emph{free structure of
    $\M$ generated by
    $R$}, denoted $F_{\M}(R)$, is the $k$-ary relation on
    $\M^{(A)}$ consisting of all tuples $(f_1,\ldots,f_k)$ for which
    there exists $g\in\M^{(R)}$ such that
    \[
        f_i=g^{\pi_i}\qquad(i\in[k]),
        \qquad
        \pi_i:R\to A,\quad
        \pi_i(a_1,\ldots,a_k)=a_i.
    \]
\end{definition}

\begin{theorem}[Characterisation of the CSinglAIP reduction]
    \label{thm:csinglaip-reduction-characterisation}
    For finite promise templates $(R,Q)$ and $(S,T)$ such that $T$ contains no constant tuple,
    \[
        \PCSP(R,Q)\leq_{\CSinglAIP}\PCSP(S,T)
        \quad\Longleftrightarrow\quad
        \omega_{\CSinglAIP}\bigl(\Pol(S,T)\bigr)
        \longrightarrow \Pol(R,Q).
    \]
\end{theorem}
\begin{proof}
    Note that the completeness of the reduction on the left is easy to show in general, so we show that the soundness of the reduction is equivalent to the property on the right.
    
    Fix a finite instance $I$ of $\PCSP(R,Q)$.  The defining property
    of the universal gadget and \Cref{lem:adjoints} give
    \[
        \Phi_S(I)\to T
        \iff
        \kCSinglAIP(\operatorname{LC}(I))\to\Pol(S,T)
        \iff
        \operatorname{LC}(I)\to
        \omega_{\text{CSinglAIP}}\bigl(\Pol(S,T)\bigr).
    \]
    The property we need is
    \cite[Lemma~C.4]{DO24_Conf_LocalConsistencyReductionConstraintSatisfactionProblemsa}
    \[
        \operatorname{LC}(I)\to\M
        \quad\iff\quad
        I\to F_{\M}(R).
    \]
    Moreover, by
    \cite[Lemma~4.4]{BBKO21_Journ_AlgebraicApproachPromiseConstraintSatisfaction},
    \[
        \M\to\Pol(R,Q)
        \quad\iff\quad
        F_{\M}(R)\to Q.
    \]
    Since $Q$ is finite, compactness shows that 
    \[
        F_{\M}(R)\to Q \quad \iff \quad (\forall I \text{ a finite $R$-instance,}\quad I\to F_{\M}(R)\implies I\to Q).
    \]
    Setting 
    $\M=\omega_{\text{CSinglAIP}}\bigl(\Pol(S,T)\bigr)$ completes the proof.
\end{proof}

We next show that the reduction from \Cref{recursive rounding reduction} can indeed be placed in this broader reduction framework.

\begin{corollary}
    \label{cor:rounding-is-csinglaip-reduction}
    Under the hypotheses of \Cref{recursive rounding reduction}, with
    $R_0=R|_{q^{-1}(0)}$,
    \[
        \PCSP(R,\NAE_c)\leq_{\CSinglAIP}\PCSP(R_0,\NAE_c).
    \]
\end{corollary}

\begin{proof}
    We construct the minion homomorphism required by
    \Cref{thm:csinglaip-reduction-characterisation}. 

    Let
    $(M,f)\in\omega_{\CSinglAIP}(\Pol(R_0,\NAE_c))^{(X)}$, let $J$ be
    the column index set of $M$, put $S=\operatorname{supp}M$, and put
    \[
        s_M(a)=\left(\sum_{x\in X}M(x,j)q(a_x)\right)_{j\in J}
        \quad\text{for }a\in\dom(R)^X.
    \]
    Extend $f$ to $\widehat f:\dom(R)^S\to[c]$ by setting
    $\widehat f(b)=f(b)$ for $b\in\dom(R_0)^S$ and
    $\widehat f(b)=1$ otherwise.  After fixing an order on each column
    index set, define
    \[
        \xi_X(M,f)(a)=
        \begin{cases}
            \chi^{(|J|)}(s_M(a)),&s_M(a)\neq0,\\
            \widehat f(a|_S),&s_M(a)=0.
        \end{cases}
    \]
    We must show that $\xi_X(M,f)$ is indeed a polymorphism. Suppose $(\bar{a},\bar{b},\bar{c}) \in R^X$. Then it is easy to see that $s_M(\bar{a})+s_M(\bar{b})+s_M(\bar{c}) = \bar{0}$.
    If some $s_M$-vector is nonzero, the $c$-rounding makes the
    three outputs nonconstant. Suppose instead that
    $s_M(\bar{a})=s_M(\bar{b})=s_M(\bar{c})=\bar{0}$. For every
    $x\in S$, the unit column $e_x$ shows that
    $q(a_x)=q(b_x)=q(c_x)=0$. Hence
    $(\bar{a}|_S,\bar{b}|_S,\bar{c}|_S)\in R_0^S$, so the three outputs
    are nonconstant since $f\in\Pol(R_0,\NAE_c)$.

    Finally, let $\pi:X\to Y$, put $S'=\pi(S)=\operatorname{supp}(M^\pi)$,
    and write $\rho=\pi|_S:S\to S'$. Since $\rho$ is surjective, for
    every $u\in\dom(R)^{S'}$ we have
    
    \[
        u\in\dom(R_0)^{S'}
        \quad\Longleftrightarrow\quad
        u\circ\rho\in\dom(R_0)^S.
    \]
    Thus $(\widehat f)^\rho=\widehat{f^\rho}$, where the right-hand side
    denotes the same constant-$1$ extension to $\dom(R)^{S'}$. Together
    with
    \[
        s_{M^\pi}(a)=s_M(a\circ\pi)
        \quad\text{and}\quad
        (a\circ\pi)|_S=(a|_{S'})\circ\rho,
    \]
    this proves that $\xi$ preserves minors.
\end{proof}

\begin{remark}
    Analogously to how the characterisation of arc-consistency reductions~\cite{DO24_Conf_LocalConsistencyReductionConstraintSatisfactionProblemsa} was lifted to a characterisation of $k$-consistency reductions~\cite{BHZ27:soda}, the same approach extends to a characterisation of the corresponding higher-level
    $\CSinglAIP$ reductions using the $\mathrm{lvl}_k$ and $\operatorname{pow}_k$ constructions of Barto, Hadek, and Zhuk~\cite{BHZ27:soda}.
\end{remark}
\section{Classification of valid partitions of \texorpdfstring{$\bm{\Z_p}$}{Zp}}\label[appendix]{appendix:valid-partitions}
Recall that we say a partition of $\Z_p$ into 3 sets $A_1, A_2, A_3$ is \emph{valid} if the only solution to $x + y + z \equiv 0 \bmod p$ with $x, y, z \in A_i$ for some $i \in [3]$ is $x = y = z = 0$. Furthermore, we only consider such partitions up to (i) multiplication by a constant and (ii) permutation of parts.

Subsets $T$ of an Abelian group for which
$t_1+t_2+t_3\neq0$ for all $t_1,t_2,t_3\in T$ are known as
\emph{$3$-independent sets}. Voloch constructed maximal
$3$-independent subsets of $\Z_p$~\cite[Section~2]{V90_Journ_CompletenessCertainPlaneArcsII},
and Sz{\H o}nyi studied their possible sizes and structure; in particular,
he proved uniqueness up to automorphism for a $3$-independent subset of
$\Z_p$ of size $(p+1)/3$ when $p\equiv2\pmod3$
~\cite[Proposition~4.2]{S91_Journ_COMBINATORIALPROBLEMSABELIANGROUPSARISINGGEOMETRY}.
\Cref{thm:partitions-seven} instead concerns the simultaneous classification of
partitions into three valid parts, with the trivial triple $(0,0,0)$ allowed
in the part containing $0$.

The proof of \Cref{thm:partitions-seven} relies on the following standard results from additive combinatorics.

\begin{theorem}[Cauchy--Davenport~\cite{Cauchy-Davenport}]
Let $p$ be prime and let $X,Y\subseteq\Z_p$ be nonempty. Then
\[
    |X+Y|\geq \min\{p,|X|+|Y|-1\}.
\]
\end{theorem}

\begin{theorem}[Vosper~\cite{Vosper}]
Let $p$ be prime and let $X,Y\subseteq\Z_p$ satisfy
$|X|,|Y|\geq2$. If
\[
    |X+Y|=|X|+|Y|-1\leq p-2,
\]
then $X$ and $Y$ are arithmetic progressions with the same nonzero
common difference.
\end{theorem}

\partitionsseven*
\begin{proof}
    It is easy to check that this is a valid partition of $\Z_p$: indeed if $x, y, z \in A_{1 + i}$ and at least one of $x, y, z$ is not 0, then the rounding operation in the following expression is applied to a nonintegral value at least once, and hence
    \[
        3i = \left \lfloor\frac{3x}{p}\right \rfloor + \left \lfloor\frac{3y}{p}\right \rfloor + \left \lfloor\frac{3z}{p}\right \rfloor < \frac{3x}{p} + \frac{3y}{p} + \frac{3z}{p}.
    \]
    Hence $ip < x + y + z$. More easily,
    \[
        3(i + 1) = \left(1 + \left\lfloor\frac{3x}{p}\right\rfloor\right) + \left(1 + \left\lfloor\frac{3y}{p}\right\rfloor\right) + \left(1 + \left\lfloor\frac{3z}{p}\right\rfloor\right) > \frac{3x}{p} + \frac{3y}{p} + \frac{3z}{p}.
    \]
    Hence $x + y + z < (i + 1)p$. So $x + y + z \not \equiv 0 \bmod p$, and this is indeed a valid partition.

    We now need to prove uniqueness up to multiplication by a constant and permutation of parts. Suppose without loss of generality that $|A_1| \leq |A_2| \leq |A_3|$. We first prove a simple fact.

    \begin{claim}\label{cl:upperBound}
        $|A_i| \leq \left \lceil \frac{p}{3} \right \rceil$.
    \end{claim}
    \begin{proof}
        Let $A = A_i$. Note that by definition any solution to $x + y + z  \equiv  0 \bmod p$ in $A$ is $x = y = z = 0$. Hence, $(A + A) \cap (-A) \subseteq \{ 0 \}$. Thus $|A + A| + |{-A}| \leq p + 1$. Note that $|{-A}| = |A|$, and furthermore by the Cauchy-Davenport theorem $|A + A| \geq \min(p, 2|A| - 1)$. Thus $\min(p, 2|A| - 1) + |A| \leq p + 1$. If the minimum on the left-hand side is $p$, then $|A| \leq 1$, which implies the desired bound. Otherwise we have $3|A| - 1 \leq p + 1$, or, rearranging, $|A| \leq \left \lfloor \frac{p + 2}{3} \right \rfloor = \left \lceil \frac{p}{3} \right \rceil$.
    \end{proof}

    Returning to the proof of uniqueness, we split into two cases: $p = 3k + 1$ and $p = 3k + 2$.

    \begin{claim}
        Uniqueness holds for $p = 3k + 1$.
    \end{claim}
    \begin{proof}
        In this case we have, by \Cref{cl:upperBound}, that without loss of generality $(|A_1|,|A_2|,|A_3|) = (k,k,k+1)$ or $(k-1,k+1,k+1)$. Note that if $|A_i|=k+1$, the inequality of \Cref{cl:upperBound} is tight, hence (i) the inequality $|A_i + A_i| + |{- A_i}| \leq p + 1$ is tight --- and hence $0 \in A_i$ --- and (ii) our application of Cauchy-Davenport is tight. Since $0$ can only be in one of the sets, we must be in the case $|A_1| = |A_2| = k$ and $|A_3| = k + 1$. Also, by Vosper's theorem, $A_3$ is an arithmetic progression. Multiplying by the inverse of the common difference of this arithmetic progression, we may assume that $A_3$ is an arithmetic progression with common difference 1, which contains 0. It cannot contain $-1$ and $1$ (or else the nontrivial solution $(-1) + 0 + 1 \equiv 0 \bmod p$ would be monochromatic). Hence it contains only one of these, and by multiplying perhaps with $-1$ we may assume without loss of generality that it contains $-1 \equiv p - 1 \bmod p$. Hence $A_3 = \{2k + 1, \ldots, 3k, 0\}$.

        At this point, we note $A_1 \cup A_2 = \{1, \ldots, 2k\}$. Note that since $k + k + (k + 1) \equiv 0 \bmod p$, we have that $k$ and $k + 1$ are not both within the same part. Assume without loss of generality $k \in A_1, k + 1 \in A_2$. We prove inductively that $\{ k - r + 1, \ldots, k \} \subseteq A_1$ and $\{k + 1, \ldots, k + r\} \subseteq A_2$ for $r = 1, \ldots, k$. The base case is already shown. Now suppose we have established this for some $r<k$. Consider that $k + 1, k + r \in A_2$ and $(k - r) + (k + 1) + (k + r) \equiv 0 \bmod p$ implies $k - r \in A_1$; and $k - r, k \in A_1$ and $(k - r) + k + (k + r + 1) \equiv 0 \bmod p$ implies $k + r + 1 \in A_2$, which completes the inductive step.

        At $r=k$, both containments are equalities by the known cardinalities of $A_1$ and $A_2$. Hence $A_1 = \{1, \ldots, k\}, A_2 = \{k + 1, \ldots, 2k\}, A_3 = \{2k+1, \ldots, 3k, 0\}$, and thus $(-A_3,-A_2,-A_1)=(\{0,\ldots,k\},
      \{k+1,\ldots,2k\},
      \{2k+1,\ldots,3k\})$. This concludes the proof of uniqueness.
    \end{proof}

    \begin{claim}
        Uniqueness holds for $p = 3k + 2$.
    \end{claim}
    \begin{proof}
        Similarly to before, we see that $|A_1| = k$ and $|A_2| = |A_3| = k + 1$. Without loss of generality, suppose $0 \not \in A_3$. Hence $(A_3 + A_3) \cap (-A_3) = \emptyset$, $|A_3 + A_3| + |{-A_3}| \leq p$, and Cauchy-Davenport holds with equality, since by disjointness  $|A_3+A_3|\leq p-|A_3|=2k+1 = |A_3|+|A_3|-1$. Hence, by Vosper's theorem, we see that $A_3$ is an arithmetic progression. 

        Suppose that $A_3$ is the arithmetic progression $\{ a + di \mid i = 0, \ldots, k\}$. Therefore $A_3 + A_3 + A_3 = \{ 3a + di \mid i = 0, \ldots, 3k \}$. Note that $0 \not \in A_3 + A_3 + A_3$, hence $3a + di$ enumerates all values $1, \ldots, 3k + 1$ as $i$ ranges from 0 to $3k$. Note furthermore that $i \mapsto 3a + di$ is a bijection on $\Z_p$, so applying it to the value not mentioned before (i.e.~$3k + 1$) yields the output not mentioned before (i.e.~$0$). In other words $3a + d(3k + 1) \equiv 0 \bmod p$. Hence $3a \equiv d \bmod p$, and multiplying by $k + 1$ we have $a \equiv d(k + 1)$. Hence $A_3 = \{ d(k+1), \ldots, d(2k + 1) \}$.

        Observe that $A_3 = -A_3$. Suppose first, for contradiction, that $0 \not \in A_2$. By the preceding argument, we see also that $A_2 = -A_2$, and hence $A_1 = -A_1$ and by the process of elimination $0 \in A_1$. But this is impossible, since then (as $k>1$) there exists some nonzero $x \in A_1$, and then $x + 0 + (-x) \equiv 0 \bmod p$ is a monochromatic nontrivial solution. Hence we find that $0 \in A_2$. Now, scale all sets by the inverse of $d$ to get to the following situation: $0 \in A_2$, $A_1 \cup A_2 = \{-k, \ldots, k\}$ and $A_3 = \{k + 1, \ldots, 2k + 1\}$.

        Note that if $A_1$ contains $x$ and $-x$ for any $x \in [k]$, by the pigeonhole principle there exists some $y \in [k]$ such that $y, -y \in A_2$, which is impossible as $0 \in A_2$. Hence for every such pair, one element belongs to $A_1$ and one to $A_2$. Assume (by negating) that $\{0, 1 \} \subseteq A_2$ and $-1 \in A_1$. We shall prove inductively that $A_1 \supseteq \{-r, \ldots, -1\}$ and $A_2 \supseteq \{0, \ldots, r\}$, for $r = 1, \ldots, k$. The base case is already proved. Now assume we have this for some $r<k$. Note that as $1, r \in A_2$ we have that $- r - 1 \in A_1$; and as $-r, -1 \in A_1$ we have that $r + 1 \in A_2$. This completes the induction.

        Hence $A_1 = \{-k, \ldots, -1\}, A_2 = \{0, \ldots, k\}, A_3 = \{ k + 1, \ldots, 2k + 1\}$. This concludes the proof of uniqueness.
    \end{proof}
    This concludes the proof of uniqueness in all cases.
\end{proof}
\partitionsfive*
\begin{proof}
    Regard a partition as a colouring
    \[
        \chi\colon \Z_5\to\{1,2,3\}.
    \]
    Since $(1,2,2),(2,4,4),(4,3,3)$, and $(3,1,1)$ are nontrivial zero-sum triples,
        \[
        \chi(1)\neq\chi(2),\quad \chi(2)\neq\chi(4),\quad \chi(4)\neq \chi(3),\quad \chi(3)\neq\chi(1).
        \]
        Thus each element in $\{\pm1\}$ cannot share a colour with each element in $\{\pm2\}$. Since only 3 colours are available, one of these sets must be monochromatic. By multiplying by $2$ and permuting the parts if necessary, 
        \[\chi(2)=\chi(3)=1.\]
        
        Since $(0,2,3)$ is a nontrivial zero-sum triple, we also have that $\chi(0)\neq1$, so $\chi(0)=3$ without loss of generality. Since $(0,1,4)$ is also a nontrivial zero-sum triple, $\chi(1)$ and $\chi(4)$ cannot both be $3$. Thus, by multiplying by $-1$ if necessary, the given partitions are the only possible options. It is easy to check that both of these are indeed valid.
\end{proof}

\section{Proof of \texorpdfstring{\Cref{thm:essentially-unary}}{essential arity theorem}}\label[appendix]{appendix:essentially-unary}
\essentiallyunary*

This section will contain three parts. First, a set of technical preliminaries. Next, a general proof of our theorem for the case $p \geq 11$. Unfortunately this proof fails for the case $p = 7$, which we treat separately in the last part.

\subsection{Technical Preliminaries}

For every prime $p\geq7$, let
\[
    h_p:\Z_p\to[3],\qquad
    h_p(x)=1+\left\lfloor\frac{3x}{p}\right\rfloor,
\]
where $x$ is represented by an integer in $\{0,\ldots,p-1\}$.
By~\Cref{thm:partitions-seven}, $h_p$ is, up to scaling its input and
permuting the colours, the unique homomorphism from
$\Eqn^*(\Z_p)$ to $\NAE_3$. 

\begin{lemma}\label{midpointclosedsets}
Let $p$ be an odd prime.
Pick disjoint sets $C, E \subseteq \mathbb{Z}_p$ such that
\[\frac{C+C}{2} \subseteq C \cup E. \]
Then either
$|C|\le |E|+1$ or $C=\Z_p\setminus E$.
\end{lemma}
\begin{proof}
The hypothesis gives $|C+C|\le|C|+|E|$.  By the
Cauchy--Davenport theorem (the conclusion is immediate if $C=\emptyset$),
$|C+C|\ge\min(p,2|C|-1)$.
If the minimum is $2|C|-1$, then $|C|\le|E|+1$.  If it is $p$, then
$p\le|C|+|E|\le p$; hence equality holds and the disjoint sets $C,E$
partition $\Z_p$.
\end{proof}

\begin{definition}
    For an odd prime $p$, we say that a set $C \subseteq \Cp$ is \emph{midpoint-closed} if, for any $r, s \in C$ with $r + s \neq 0$, we have $\frac{r + s}{2} \in C$.
\end{definition}

\begin{corollary}\label{midpointclosedsetscor}
    Let $p$ be an odd prime. If $C \subseteq \Cp$ is midpoint closed and admits elements $a, b \in \Cp$ with $a \neq \pm b$, then $C = \Cp$.
\end{corollary}
\begin{proof}
Pick such $a,b\in C$.
By assumption, $a,b,\frac{a+b}{2}$ are distinct and are all in $C$, hence $|C|\geq 3$.
The conclusion follows by \Cref{midpointclosedsets} with $E=\{0\}$.
\end{proof}

\begin{lemma}\label{intervaldilates}
    Let $p\geq 7$ prime and let $k =\left\lfloor\frac{p}{3}\right\rfloor$, $S = \{1,\dots,k\}$, $M = \Cp \setminus (S \cup -S)$. Equivalently, $S=h_p^{-1}(1)\setminus\{0\},M=h_p^{-1}(2),-S=h_p^{-1}(3)$. Then for $d,c\in \Cp$:
    \begin{lemenumerate}
        \item\label{intervaldilates(1)} If $dS\cap(-S) = \emptyset$, then $d\in \{1,2,\frac{p+1}{2}\}$.
        \item\label{intervaldilates(2)} If $cM\cap M=\emptyset$, then (i) $c \in \{ \pm 2, \pm \frac{1}{2} \}$ if $p \neq 13$, or (ii) $c \in \{ \pm 2, \pm \frac{1}{2}, \pm 5\}$ if $p = 13$. 
        \item\label{intervaldilates(3)} If $cM=M$, then $c \in \{\pm1\}$.
    \end{lemenumerate}
\end{lemma}
We prove each point separately.
\begin{proof}[Proof of \Cref{intervaldilates(1)}]
    Suppose $d\not\in \{1,2,\frac{p+1}{2}\}$. In the following, $d$ is taken to be its representative within $1, \ldots, p - 1$.

    \begin{description}
    \item[Case $\bm{3\leq d < p/3}$.]
    Then $ \lfloor p/d \rfloor \in S$, and $-d\lfloor p/d \rfloor \in S$, contradicting the assumption $dS \cap (-S) = \emptyset$.
    \item[Case $\bm{p/3<d\leq (p-1)/2}$.]
    Then $2,-2d \in S$, likewise leading to a contradiction.
    
    \item[Case $\bm{(p+3)/2\leq d<2p/3}$.] This interval is empty for $p=7$, so we may assume $p\geq 11$. Let $a = 2d-p$ and $n=2\lfloor k/a\rfloor+1$.
    By a simple computation we have $n \in S$ and $p-2k \leq na< p$. Since $n$ is odd and $d=(p+a)/2$,
    \[-nd = \frac{-np - na}{2} = -\frac{n + 1}{2} p + \frac{p - na}{2} \equiv \frac{p-na}{2} \pmod{p}\]
    Using the inequality for $na$, we finally have $0<-nd\leq k$ and hence $-nd \in S$.
    \item[Case $\bm{2p/3<d<p}$.]
    Then $1,-d \in S$.\qedhere
\end{description}
\end{proof}
\begin{proof}[Proof of \Cref{intervaldilates(3)}.] If $cM=M$, then squaring and taking the difference we have
\[
(c^2-1)\sum_{x\in M}x^2=0.
\]
The total sum of squares in $\Z_p$ is $\frac{(p - 1) p (2p - 1)}{6} \equiv 0$, hence as $\Cp \setminus M = \{\pm1,\dots,\pm k\}$ we have
\[
\sum_{x\in M}x^2
\equiv-2\sum_{j=1}^k j^2
\equiv-\frac{k(k+1)(2k+1)}{3}\not \equiv0\pmod p.
\]
Thus $c^2=1$, proving item~(3).
\end{proof}
\begin{proof}[Proof of \Cref{intervaldilates(2)}.]
Since $M=-M$, we have $(-c)M=cM$. Thus, after replacing $c$ by
$-c$ if necessary, we may assume
$1\leq c\leq\frac{p-1}{2}$.
We first deal directly with $p=7,11,13$. Multiplying 
$M$ by each possible value of $c$ in the specified interval yields:
\[
\begin{array}{c|c|c}
p&M&
\left\{1\leq c\leq\frac{p-1}{2}:cM\cap M=\emptyset\right\}\\ \hline
7&\{3,4\}&\{2,3\}\\
11&\{4,5,6,7\}&\{2,5\}\\
13&\{5,6,7,8\}&\{2,5,6\}.
\end{array}
\]
Noting that $3 \equiv -\frac{1}{2} \bmod 7$, $5 \equiv -\frac{1}{2} \bmod 11$ and $6 \equiv -\frac{1}{2} \bmod 13$ completes the proof for this case.

Now, let $p\geq 17$ and $cM\cap M=\emptyset$, so $cM \subseteq S\cup(-S)$. Write $m=|M|$. For $i=1,\ldots,m$, let $x_i$ be the representative of $c(k+i)$ in the interval $[-k,k]$. Since $x_{i+1}-x_i \in [-2k,2k]$ for $i=1,\ldots,m-1$, we have
$x_{i+1}-x_i \in \{c,c-p\}$.
We now split in cases depending on whether $c \leq m$ or $c \geq m + 1$.

\begin{description}
\item[Case $\bm{c\leq m}$.] Since $p-m=2k+1$, the value $c-p$ is strictly smaller than $-2k$, so we must have $x_{i + 1} - x_i = c$ for all $i = 1, \ldots, m-1$.
Hence by a telescoping sum, $(m - 1)c = x_m - x_1 \leq 2k$; since $m \in \{k, k + 1\}$, this implies $c \leq 2$. Since of course $M = 1M$, we must have $c = 2$.

\item[Case $\bm{c\geq m+1}$.] As $m\geq k$, we have $c>k$, and since $c\leq\frac{p-1}{2}$, we also have $c-p<-k$. Thus, for the $x_i$'s to stay in $[-k,k]$, the increments $x_{i+1}-x_i$ must alternate between $c$ and $c-p$.

Put \(q=p-2c\), so \(q\) is a positive odd integer. Since \(m=p-1-2k\) is even, put \(L=\frac m2-1\). The two parity subsequences have common difference \(-q\), and hence $x_{m-1}=x_1-Lq$ and $x_m=x_2-Lq$.
As all the \(x_i\) lie in \([-k,k]\), we obtain
\[
\begin{aligned}
2k
&\geq \max\{x_1,x_2\}-\min\{x_{m-1},x_m\}\\
&=|x_2-x_1|+Lq\\
&\geq c+Lq,
\end{aligned}
\]
where the final inequality uses \(x_2-x_1\in\{c,c-p\}\) and \(c\leq p-c\).
Now \(c=(p-q)/2\), so $p + (m - 3)q \leq 4k$.
Since \(p=3m\pm1\) and $2k + m = p - 1$, we have $2k = 2m - 1 \pm 1$. Substituting into $p + (m- 3)q \leq 4k$, we get
\[
3m \pm 1 + (m - 3)q \leq 4m - 2 \pm 2,
\]
where the $\pm$ signs are the same on both sides. This implies $(m - 3)q \leq m - 2 \pm 1$.

As \(p\ge17\), we have \(m\ge6\). Since \(q\) is a positive odd integer,
the inequality forces \(q=1\). Therefore
\[
c=\frac{p-q}{2}=\frac{p-1}{2}.\qedhere
\]
\end{description}
\end{proof}
\subsection{Restricted binary polymorphisms for \texorpdfstring{$\bm{p\geq 11}$}{p >= 11}}
    Throughout this subsection, let $p\geq 11$ be an odd prime, and let $k= \lfloor p/3 \rfloor$.
    Let $\tau$ be the permutation swapping 2 and 3.
    Below we let $x \oplus y$ denote addition modulo 2; of course $\tau^{x \oplus y} = \tau^x \circ \tau^y$.
    Let $f \in \Pol^{(2)}(\Eqn^*(\Z_p),\NAE_3)|_\Cp$.
    Choose $\widetilde f\in\Pol^{(2)}(\Eqn^*(\Z_p),\NAE_3)$ whose restriction to $\Cp$ is $f$.
    After simultaneously scaling both inputs and permuting the colours, we may
    assume by the classification of valid partitions that $\widetilde f(t,t)=h_p(t)$ for every $t\in\Z_p$.
    
    For each \(r\in\Cp\), the map \(t\mapsto \widetilde f(t,rt)\) is a
    homomorphism \(\Eqn^*(\Z_p)\to\NAE_3\) sending \(0\) to \(1\). Hence, by
    the classification of valid partitions, for every $r \in \Cp$, there exist
    \[
    P_r\in(\Cp)^2,\qquad \epsilon(r)\in\{0,1\},
    \]
    where \(P_r\) has slope \(r\) (meaning $P_r = (x, rx)$ for some $x \in \Cp$), such that
    \[
    f(tP_r)=\tau^{\epsilon(r)}h_p(t)\qquad(t\in\Cp).
    \]
Note that since $f(x,x)=h_p(x)$ for every $x \in \Cp$, we have $P_1 = (1, 1)$ and $\epsilon(1) = 0$.

We want to show that $f$ is essentially unary. Equivalently:
    \begin{enumerate}
        \item $\epsilon(r)=0$ for all $r$ and
        \item either $P_r = (1,r)$ for all $r$ or $P_r = (r^{-1},1)$ for all $r$.
    \end{enumerate} 
First we need to prove some more lemmas. Since $p=13$ is in some sense special, it will be useful to define
\[D \coloneq \begin{cases}
\D, &\text{if $p\neq13$,}\\
\Ds, &\text{if $p=13$.}
\end{cases}\]
\begin{lemma}\label{coefficients}
Let $p\geq 11$ prime and $r,s \in \Cp$. Define $\alpha: \pm D \to \{0,1\}$, $\alpha(\pm1)=1, \alpha(x)=0$ otherwise.
\begin{lemenumerate}
    \item\label{coefficients(1)} Suppose $P_r+P_s \in (\Cp)^2$, and let $q,d\in \Cp$ be such that $P_r+P_s=dP_q$. Then
    $d \in \{1,2,\frac{1}{2}\}$
    and
    \[\epsilon(r)=\epsilon(s)\eqcolon e \implies \epsilon(q)=e\oplus \alpha(d).\]
    \item\label{coefficients(2)} Suppose $P_r-P_s \in (\Cp)^2$, and let $q,c\in \Cp$ be such that $P_r-P_s=cP_q$. Then 
    \[\epsilon(r)=\epsilon(s)\eqcolon e \implies c \in \pm D \text{ and }\epsilon(q)=e\oplus \alpha(c).\]
\end{lemenumerate}
    
\end{lemma}
\begin{proof}[Proof of \Cref{coefficients(1)}]
    Since $P_r+P_s -dP_q=(0,0)$, we have:
    \[(f(tP_r),f(tP_s),f(-dtP_q))=(\tau^{\epsilon(r)}h_p(t),\tau^{\epsilon(s)}h_p(t),\tau^{\epsilon(q)}h_p(-dt))\in \NAE_3\]
    for all $t \in \Cp$. In particular, taking $t\in S$, we get by \Cref{intervaldilates(1)} that $d\in \{1,2,\frac{p+1}{2}\}$.
    
    If $\epsilon(r)=\epsilon(s)\eqcolon e$, we get
    $\tau^{e\oplus\epsilon(q)}h_p(-dt)\neq h_p(t)$
    for all $t \in \Cp$.
    Now, consider possible values of $(d,\epsilon(q))$. $(2,e)$ is valid from the zero-sum triple
$(t,t,-2t)$, and $(1/2,e)$ from $(-t/2,-t/2,t)$.  Also
$\tau h_p(-t)\ne h_p(t)$ for every $t\ne0$, as is immediate from the three
    intervals defining $h_p$.  The remaining tuples $(1,e)$, $(2,1\oplus e)$, and $(\frac{1}{2},1 \oplus e)$ fail respectively by
taking $t$ in $h_p^{-1}(2)$, $t=\frac{1}{2}$, and $t=-1$.
\end{proof}
\begin{proof}[Proof of \Cref{coefficients(2)}]
    Assume $\epsilon(r)=\epsilon(s)\eqcolon e$. Since $P_r-P_s -cP_q=(0,0)$, we have:
    \[(f(tP_r),f(-tP_s),f(-ctP_q))=(\tau^eh_p(t),\tau^eh_p(-t),\tau^{\epsilon(q)}h_p(-ct)) \in \NAE_3\]
    for all $t$.
    Taking $t\in M$, we get that $h_p(t)=h_p(-t)$, hence (replacing $t$ by $-t$) $\tau^{e\oplus\epsilon(q)}h_p(ct)\neq h_p(t)$.
\begin{itemize}
    \item If $e \oplus \epsilon(q) = 0$, the tuple implies $cM\cap M=\emptyset$, so \Cref{intervaldilates(2)}
  gives $c\in\{\pm2,\pm1/2\}$ for $p\neq 13$ and $c\in\{\pm2,\pm1/2,\pm5\}$ for $p=13$, i.e.~$c \in \pm D$.
    \item If $e\oplus \epsilon(q) =1$, it implies that the symmetric set $cM=-cM$ avoids
  $h_p^{-1}(3)=-S$.  If $cM$ met $S$, symmetry would make it meet $-S$ as
  well. Thus, $cM$ contains neither $0$ nor any point of $S\cup(-S)$, and therefore
  $cM\subseteq M$.  Equal cardinalities give $cM=M$, and \Cref{intervaldilates(3)} gives
  $c=\pm1 \in \pm D$.\qedhere
\end{itemize}
\end{proof}
\begin{lemma}\label{collision}
Let $p\geq 11$ and define for $a\in D$
\[\phi_a(t)\coloneq \tau^{\alpha(a)}h_p(at).\]
 For every $a,b\in D$, there is $t\in\Cp$ such that
\[
\phi_a(t)=\phi_b(t)=h_p(-2t).
\]
\end{lemma}

\begin{proof}
This follows by case exhaustion. Write $p=6 n\pm1$. We calculate:
\[
\begin{array}{c|c|c}
\text{Condition on }a,b&t&
\phi_a(t)=\phi_b(t)=h_p(-2t)\\ \hline
a,b\in\{2,\frac{1}{2}\}&2n-1&2\\
a=b=1&\frac{1}{2}&3\\
\{a,b\}=\{1,2\}&-k&2\\
\{a,b\}=\{1,\frac{1}{2}\}&4n-1&3
\end{array}
\]
It remains to consider $p=13$ and the pairs involving $5$. Since
$\frac{1}{2}=7$ in $\Z_{13}$, we can calculate:
\[
\begin{array}{c|c|c}
\text{condition on }a,b&t&
\phi_a(t)=\phi_b(t)=h_p(-2t)\\ \hline
5\in\{a,b\}\subseteq\{1,\frac{1}{2},5\}&\frac{1}{2}&3\\
\{a,b\}=\{2,5\}&4&2
\end{array}
\qedhere
\]

\end{proof}

\begin{lemma}\label{starlemma}
    Assume $p\ge11$.  If $\epsilon(r)=\epsilon(s)\eqcolon e$, then for $P=P_r=(P_x,P_y)$ and $Q=P_s=(Q_x,Q_y)$ we have
$$
P_x=\pm Q_x
\qquad\text{or}\qquad
P_y=\pm Q_y.
$$
\end{lemma}

\begin{proof}
Suppose that neither coordinate of $P=P_r$ agrees up to sign with
the corresponding coordinate of $Q=P_s$. This means that the sums $P + Q$ and $P - Q$ have no zero coefficients. Hence we may apply \Cref{coefficients} to deduce that $P + Q = dR$ and $P - Q = \pm c T$ for some $R = P_v, T = P_u, d \in \D, c \in D$. Furthermore, $\epsilon(v) \oplus e = \alpha(d), \epsilon(u) \oplus e = \alpha(c)$.

Let $dR + cT = 2X$; note $X \in \{P, Q\}$.
Apply \Cref{collision} to $c$ and $d$ to yield $t$.
After applying $\tau^e$ to
all three colours, the zero-sum triple
$tdR+tcT-2tX=0$
is mapped to
$\bigl(\phi_d(t),\phi_c(t),h_p(-2t)\bigr)$ by $f$,
which is monochromatic. This is a contradiction.
\end{proof}

\begin{theorem}
Let $p\geq11$ prime and $f,\epsilon(r),P_r$ be as described at the start of this subsection. Then
\begin{enumerate}
\item $\epsilon(r)=0$ for every $r\in\Cp$, and
\item either $P_r=(1,r)$ for every $r\in\Cp$, or
      $P_r=(r^{-1},1)$ for every $r\in\Cp$.
\end{enumerate}
Equivalently, $f$ is essentially unary.
\end{theorem}

We will devote the remainder of the section to proving this theorem. Let us first give a schematic overview of the proof. We begin by creating a family of (two) graphs derived from the polymorphisms we are studying, with edges induced by the $P_r$'s. We deduce that each such graph is bipartite and its edges intersect pairwise; hence it is a star, which enforces consistency (in some sense) between the edges within each graph. We then prove in stages that in fact only one graph within our family has any edges at all. The final step is to decode one such star into a coordinate unambiguously.

\paragraph{Stage 1: Graph representation.}

Let $\Cp/\{\pm1\}$ be $\Cp$ quotiented by sign, writing $[x]$ for
the equivalence class $\{x,-x\}$. For each $e\in\{0,1\}$, define the
bipartite graph $G_e$ with vertex set two copies of $\Cp/\{\pm1\}$
and edge set
\[
\left\{([x],[y])\mid
  \exists r\in\epsilon^{-1}(e):P_r=(x,y)\right\}.
\]
Note that every $P_r$ induces an edge in precisely one graph $G_e$, namely in $G_{\epsilon(r)}$.
By \Cref{starlemma}, any two edges of $G_e$ are incident. A
pairwise-intersecting family of edges in a bipartite graph has a
common vertex --- any intersecting family of edges is either a star or a triangle, and the triangle is not bipartite.
Thus $G_e$ is a
star, together with possibly some isolated vertices.

Choose $e\in\{0,1\}$ such that, if we take $A = \epsilon^{-1}(e), B = \epsilon^{-1}(e \oplus 1)$, then we have $|A| \geq |B|$. In particular, we then have $|A| \geq \frac{p - 1}{2} > 0$. Consider the graph $G_e$, and assume without loss of generality that the star has its central vertex on the left side of $G_e$. (If it is on the right side, this can be fixed by swapping coordinates of $f$, which swaps coordinates of every $P_r$ and thus also swaps the sides of $G_e$.) Letting $[\gamma]$ be this central vertex, we see that every $P_r$ with $r \in A$ is of the form $\pm \gamma(1, r)$. For each $r \in A$, choose $\delta_r \in \{\pm 1\}$ such that $P_r = \delta_r \gamma(1, r)$.

\paragraph{Stage 2: Characterising $\bm{B}$.}
Set $T=\{r\in B\mid P_r = \pm2\gamma(1, r) \}$. Our proof will first show that this set must be small, then it will show that in fact $T = B$.
For $r \in T$ define $\delta_r \in \{ \pm 1 \}$ so that $P_r = 2 \delta_r \gamma(1, r)$.
We shall use the following simple consequence of \Cref{coefficients}.

\begin{corollary}\label{cor mix}
    Suppose $\epsilon(q) = \epsilon(s)$ and $P_q \pm P_s = c P_m \in (\Cp)^2$. Furthermore, suppose $\epsilon(q) \neq \epsilon(m)$. Then $c = \pm 1$.
\end{corollary}

\begin{claim}
$T$ intersects at most one set of the form $\{ \pm x \}$, and hence $|T| \leq 2$.
\end{claim}

\begin{proof}
Suppose for contradiction that there exist $r,r'\in T$ with
$r\neq\pm r'$. These induce lines $P_r$ and $P_{r'}$ which in turn induce the edges $([2\gamma], [2\gamma r])$ and $([2\gamma], [2 \gamma r'])$ within $G_{e \oplus 1}$. These edges share a left endpoint (but not a right endpoint) --- hence $G_{e \oplus 1}$ is a star with its central vertex at the shared vertex $[2 \gamma]$. But then every $r \in B$ satisfies $P_r = \pm 2 \gamma (1, r)$, and thus $B = T$.

We now show $T$ is midpoint-closed: consider $q, s \in T$ with $q + s \neq 0$, and take $m = \frac{q + s}{2} \in \Cp$. Observe
\[
P_q + \delta_q \delta_s P_s = 4\delta_q \gamma(1, m) = cP_m,
\]
for some constant $c$ (since at this point we already know that $P_m$ is either $\pm \gamma(1, m)$ or $\pm 2\gamma(1, m)$). Assume for contradiction that $m \in A$, hence $\epsilon(m) = e$ and $P_m = \pm \gamma(1, m)$. Then we are exactly in the situation from \Cref{cor mix} and we deduce $c = \pm 1$. This implies $\pm \gamma(1, m) = P_m = \pm 4 \gamma(1, m)$, hence $\pm 1 = \pm 4$, impossible in $\Cp$ for $p \geq 11$.

Thus $T$ is midpoint-closed. Since it contains $r,r'$ with
$r\neq\pm r'$, \Cref{midpointclosedsetscor} gives
$T=\Cp$, contradicting $A\neq\emptyset$. Hence $T$ intersects at most 1 set of the form $\{ \pm x \}$.
\end{proof}

\begin{claim}
$\displaystyle \frac{A+A}{2}\subseteq A\cup T\cup\{0\}$.
\end{claim}
\begin{proof}
Let $r,s\in A$ with $r+s\neq0$, and put
$m=\frac{r+s}{2}\in\Cp$.
Then
\[
P_r+\delta_r\delta_sP_s
=
2\delta_r\gamma(1,m)
=
cP_m
\]
for some constant $c$.
If $\epsilon(m)=e$, then $m\in A$. Otherwise, \Cref{cor mix} gives
$c=\pm1$, hence $P_m = \pm 2 \gamma(1, m)$ and $m \in T$. 
Allowing also the case $r+s=0$, we deduce $\frac{A + A}{2} \subseteq A \cup T \cup \{ 0 \}$.
\end{proof}

\begin{claim}
$B = T$.
\end{claim}
\begin{proof}
Apply \Cref{midpointclosedsets} with $C=A$ and $E=T\cup\{0\}$. We deduce that either $|A| \leq |T| + 2$ or $A \cup T = \Cp$. But the first alternative is impossible: $|T| \leq 2$ and thus $|T| + 2 \leq 4$, yet $|A| \geq \frac{p -1}{2} \geq 5$. Thus $A$ and $T$ partition $\Cp$. Since also $A$ and $B$ partition $\Cp$, it follows $B = T$.
\end{proof}

\paragraph{Stage 3: Eliminating $\bm{B}$.}
We next show that $B=\emptyset$. First, partition $A$ in terms of the signs of the first coordinate, i.e.~for $\delta\in\{\pm1\}$, define
\[
A_\delta
=
\{r\in A\mid P_r=\delta\gamma(1,r)\}
=
\{r\in A\mid\delta_r=\delta\}.
\]
\begin{claim}
$B = \emptyset$. Hence $\Cp = A = A_+ \cup A_-$.
\end{claim}
\begin{proof}

Suppose otherwise, and choose
$q\in T=B$. Write $\delta=\delta_q$, so that $P_q = 2 \delta \gamma(1, q)$.

We first claim that $|A_\delta| \geq 2$.
Otherwise,
$|A_{-\delta}| \geq |A|-1 = p-2-|T| \geq p-4 >\frac p2$.
Hence, by Cauchy-Davenport, $A_{-\delta}+A_{-\delta}=\Z_p$, so there exist
$r,s\in A_{-\delta}$ such that $r+s=2q$. But then
\[
P_r+P_s
=
-2\delta\gamma(1,q)
=
-P_q,
\]
contradicting \Cref{coefficients(1)}. We may therefore choose $r\in A_\delta$ with $r\neq-2q$. Put
$u=\frac{2q+r}{3}\in\Cp$.
Then
\[
P_q+P_r
=
3\delta\gamma(1,u)
=
dP_u.
\]
If $u\in A$, comparison of first coordinates gives $d=\pm3$,
whereas if $u\in B=T$, it gives $d=\pm\frac32$. This contradicts \Cref{coefficients(1)}: $\pm 3$ and $\pm \frac{3}{2}$ are not equivalent modulo $p \geq 11$ to $1, 2, \frac{1}{2}$.
Therefore $B=\emptyset$.
\end{proof}

\paragraph{Stage 4: Eliminating one of $\bm{A_+}$ and $\bm{A_-}$ and decoding.} We start by showing that $A_+$ and $A_-$ are midpoint closed.

\begin{claim}
    For $\delta = \pm 1$, $A_\delta$ is midpoint-closed.
\end{claim}
\begin{proof}
Take $r,s\in A_\delta$ with
$r+s\neq0$. Setting $m = \frac{r+s}{2}$, we have
\[
P_r+P_s
=
2\delta\gamma(1,m)
=
dP_m.
\]
Since $B=\emptyset$, we have $m\in A$, and $P_m = \pm \gamma(1, m)$. Thus $2\delta \gamma(1, m) = \pm d \gamma(1, m)$ and $d = \pm 2$. Conversely, by \Cref{coefficients(1)} we have $d \in \{1, 2, \frac{1}{2}\}$, and as $p \geq 11$ this implies $d = 2$. Thus $P_m = \delta \gamma(1, m)$, and $m \in A_\delta$.
\end{proof}

At least one of $A_+$ and $A_-$ contains two elements which are not
equal up to sign. Indeed, otherwise each $A_\delta$ would have size at most 2, giving $|\Cp|\leq 2+2=4$, a contradiction. 
By \Cref{midpointclosedsetscor}, this $A_\delta$ equals $\Cp$. Thus we deduce the following fact.
\begin{fact}
For some $\delta \in \{ \pm 1 \}$, we have $P_r = \delta \gamma(1, r)$ for all $r \in \Cp$, or $P_r = \delta\gamma(r^{-1}, 1)$ for all $r \in\Cp$. Furthermore, $\epsilon(r) = 0$ for all $r \in \Cp$.
\end{fact}
\begin{proof}
Since $P_1=(1,1)$ and $\epsilon(1)=0$, we obtain $\delta \gamma = 1$ and $e = 0$. Thus, if we did not interchange coordinates earlier, we have $P_r = (1, r)$ and $\epsilon(r) = 0$ for all $r \in \Cp$. Conversely, if we did interchange coordinates we have $P_r = (r^{-1}, 1)$ and again $\epsilon(r) = 0$ for all $r \in \Cp$.
\end{proof}

Let us now briefly argue this implies that the operations are essentially unary. If $P_r = \delta \gamma(1, r)$,
for $(x,y)\in(\Cp)^2$, taking $r=y/x$ gives
$(x,y)=xP_r$ and hence $f(x, y) = h_p(x)$. If $P_r = \delta \gamma(r^{-1}, 1)$ again with $r = y / x$ we have $(x, y) = yP_r$ and hence $f(x, y) = h_p(y)$.
Thus $f$ is essentially unary.

\subsection{Restricted binary polymorphisms for \texorpdfstring{$\bm{p=7}$}{p=7}}

Much of the setup of this section will be similar to the previous one --- we redo it to refresh the reader's memory. Let $\tau$ be as in the preceding subsection, and let
$f\in\Pol^{(2)}(\Eqn^*(\Z_7),\NAE_3)|_{\Z_7^*}$.
Choose
$\widetilde f\in\Pol^{(2)}(\Eqn^*(\Z_7),\NAE_3)$
whose restriction to $(\Z_7^*)^2$ is $f$. As in the preceding
subsection, after simultaneously scaling both inputs and permuting the
colours, we may assume that
\[
\widetilde f(t,t)=h_7(t)
\qquad(t\in\Z_7).
\]

For every $r\in\Z_7^*$, the map $t \mapsto \widetilde f(t, rt)$
is a homomorphism from $\Eqn^*(\Z_7)$ to $\NAE_3$ which sends $0$ to
$1$. Hence, by the classification of valid partitions, there exist
\[
P_r\in(\Z_7^*)^2,
\qquad
\epsilon(r)\in\{0,1\},
\]
where $P_r$ has slope $r$, such that
\[
f(tP_r)=\tau^{\epsilon(r)}h_7(t)
\qquad(t\in\Z_7^*).
\]
Write $P_r=a_r^{-1}(1,r)$ for some $a_r\in\Z_7^*$. Equivalently,
\[
f(t,rt)=\tau^{\epsilon(r)}h_7(a_rt)
\qquad(t\in\Z_7^*).
\]
The diagonal normalisation allows us to take
\[
a_1=1,
\qquad
\epsilon(1)=0.
\]
Thus, on each of the other five lines through the origin, the
restriction of $f$ is one of the twelve distinct maps
\[
t\longmapsto\tau^e h_7(at),
\qquad
(a,e)\in\Z_7^*\times\{0,1\}.
\]

\begin{claim}\label{cl:p7-finite-check}
Suppose that, for every $r\in\Z_7^*$, we choose
$(a_r,e_r)\in\Z_7^*\times\{0,1\}$, with $(a_1,e_1)=(1,0)$, and
define
\[
g(t,rt)=\tau^{e_r}h_7(a_rt)
\qquad(r,t\in\Z_7^*).
\]
If
$g:\Eqn^*(\Z_7)^2|_{(\Z_7^*)^2}\to\NAE_3$,
then exactly one of the following holds:
\begin{enumerate}
    \item $(a_r,e_r)=(1,0)$ for every $r\in\Z_7^*$;
    \item $(a_r,e_r)=(r,0)$ for every $r\in\Z_7^*$.
\end{enumerate}
\end{claim}

\begin{proof}
We prove the claim by exhaustion using the Python script in
\Cref{lst:p7-verifier}, which was implemented with GPT-5.6 Sol (cf. AI methodology).
Lines~3--13 of \Cref{lst:p7-verifier} construct the set
\[
\mathtt{TYPES}=\Z_7^*\times\{0,1\}
\]
and, for each $(a,e)\in\mathtt{TYPES}$, store the values of the map
\[
t\longmapsto\tau^e h_7(at)
\]
in \texttt{RAY\_VALUES[(a,e)]}. Line~13 verifies that these twelve
maps are distinct.

Lines~14--26 identify a point $u=(x,y)\in(\Z_7^*)^2$ with its
ray coordinates $(r,t)=(y/x,x)$ and construct the set
\[
\bigl\{(u_1,u_2,u_3)\in((\Z_7^*)^2)^3
  \mathrel{\big|}u_1+u_2+u_3=(0,0)\bigr\}
\]
up to permutation. There are $168$ such three-element multisets, where
repetitions of points are allowed; this count is printed on line~42.
For a tuple
\[
\mathtt{choice}=((a_1,e_1),\ldots,(a_6,e_6)),
\]
lines~28--33 evaluate $g(t,rt)$ using
\texttt{RAY\_VALUES[(a\_r,e\_r)]} and test that all $168$ triples are
mapped nonmonochromatically.

There are $12^5=248\,832$ choices after fixing $(a_1,e_1)=(1,0)$.
Lines~35--39 exhaustively test all of them. The assertion on line~41
checks that the only two which pass all $168$ constraints are
\[
\begin{aligned}
((a_r,e_r))_{r=1}^6
&=((1,0),(1,0),(1,0),(1,0),(1,0),(1,0)),\\
((a_r,e_r))_{r=1}^6
&=((1,0),(2,0),(3,0),(4,0),(5,0),(6,0)),
\end{aligned}
\]
which are precisely the two choices in the statement.
\end{proof}

\begin{theorem}
Every function in
$\Pol^{(2)}(\Eqn^*(\Z_7),\NAE_3)|_{\Z_7^*}$ is essentially
unary.
\end{theorem}

\begin{proof}
Apply \Cref{cl:p7-finite-check} to the parameters above. In the first
case, for $(x,y)\in(\Z_7^*)^2$, put $r=y/x$ and $t=x$.
Then
\[
f(x,y)=f(t,rt)=h_7(t)=h_7(x).
\]
In the second case, the same substitution gives
\[
f(x,y)=f(t,rt)=h_7(rt)=h_7(y).
\]
Thus the normalised function $f$ depends on only one coordinate.
Undoing the simultaneous scaling of the inputs and the permutation
of the colours preserves this property, so the original function is
essentially unary as well.
\end{proof}

\clearpage
\begin{lstlisting}[
    caption={Complete dependency-free Python verification of
    \Cref{cl:p7-finite-check}.},
    label={lst:p7-verifier}
]
from itertools import product
P = 7; U = range(1, P)
# The twelve possible maps t -> tau^e(h_7(a*t)).
def h_P(x):
    return 1 + (3 * (x % P)) // P

TAU = (0, 1, 3, 2)
TYPES = list(product(U, (0, 1)))
RAY_VALUES = {
    (a, e): tuple((TAU[h_P(a * t)] if e else h_P(a * t)) for t in U)
    for a, e in TYPES
}
assert len(set(RAY_VALUES.values())) == (P-1)*2
# The unordered triples in (Z_7^*)^2 which sum to (0,0).
def ray_coordinates(u):
    x, y = u
    return y * pow(x, -1, P) % P, x

POINTS = set(product(U, repeat=2))
TRIPLES = set()
for u in POINTS:
    for v in POINTS:
        w = ((-u[0] - v[0]) % P, (-u[1] - v[1]) % P)
        if w in POINTS:
            TRIPLES.add(tuple(sorted((u, v, w))))
CONSTRAINTS = [tuple(ray_coordinates(u) for u in c) for c in TRIPLES]
# Test whether all triples are mapped nonmonochromatically.
def colour(choice, point):
    r, t = point
    return RAY_VALUES[choice[r - 1]][t - 1]

def is_valid(choice):
    return all(len({colour(choice, u) for u in c}) > 1 for c in CONSTRAINTS)
# Fix (a_1,e_1)=(1,0) and exhaust the remaining 12^5 choices.
SOLUTIONS = []
for tail in product(TYPES, repeat=P-2):
    choice = ((1, 0),) + tail
    if is_valid(choice):
        SOLUTIONS.append(choice)
EXPECTED = [((1, 0),) * 6, tuple((r, 0) for r in U)]
assert SOLUTIONS == EXPECTED
print("constraints:", len(CONSTRAINTS), "search space:", len(TYPES)**5)
print(*SOLUTIONS, sep="\n")
\end{lstlisting}

\paragraph{Acknowledgements.} 
Nakajima thanks Amey Bhangale for pointing out the connection between Abelian embeddings and reconfigurability, and Demian Banakh for useful discussions. 

\paragraph{AI methodology.}
LLMs were used for proving the hardness side of our results. Nakajima used Opus 4.8, Opus 5, Sonnet 5, and DeepSeek interactively to help understand the combinatorics of our problems (e.g.~writing SAT code, searching for useful properties). Verwimp also used GPT-5.6 Sol to assist in proving \Cref{thm:essentially-unary}.
Following an iterative exchange and prompts by Verwimp to use~\Cref{thm:partitions-seven}, the model found the proof for primes $p\geq 11$ presented in \Cref{appendix:essentially-unary}.
ChatGPT was also used to reimplement the self-contained $p=7$ verification in \Cref{lst:p7-verifier}. The models above also helped with a literature search. ChatGPT was also used to improve the presentation of the paper, including the generation of the figure; and to look for typos and minor errors.
The authors take full responsibility
for the final text.

\phantomsection
\addcontentsline{toc}{section}{References}
{\small
\bibliographystyle{alphaurl}
\bibliography{bibliography}
}

\end{document}